\documentclass{revtex4}

\usepackage{amsmath}
\usepackage{amssymb}
\usepackage{amsfonts}
\usepackage{bm}
\usepackage[dvips,xdvi]{graphicx}
\usepackage{color}

\begin{document}


%
%
%

\title{Concentration Dependence of
Rheological Properties of Telechelic Associative Polymer Solutions}

\author{Takashi Uneyama}
\email{uneyama@se.kanazawa-u.ac.jp}
\affiliation{%
  JST-CREST and Institute for Chemical Research, Kyoto University, Gokasho,
  Uji 611-0011, Japan
}%
\affiliation{%
  School of Natural System, College of Science and Engineering,
  Kanazawa University, Kakuma, Kanazawa 920-1192,
  Japan
}%

\author{Shinya Suzuki}
\affiliation{%
  Institute for Chemical Research, Kyoto University, Gokasho,
  Uji 611-0011, Japan
}%
\affiliation{%
Lintec corporation, \\
5-14-42 Nishikicho, Warabi, Saitama 335-0005, Japan
}%

\author{Hiroshi Watanabe}
\affiliation{%
  Institute for Chemical Research, Kyoto University, Gokasho,
  Uji 611-0011, Japan
}%

\date{\today}


\begin{abstract}
We consider concentration dependence of rheological properties of
associative telechelic polymer solutions. Experimental results for model
telechelic polymer solutions show rather strong concentration dependence
of rheological properties.
For solutions with relatively high concentrations, linear viscoelasticity
deviates from the single Maxwell behavior. The concentration dependence of
characteristic relaxation time and moduli is different in high and low
concentration cases.
These results suggest that there are two different concentration regimes.
We expect that
densely connected (well percolated) networks are formed in
 high-concentration solutions, whereas sparsely connected 
(weakly percolated) networks are formed in low-concentration solutions.
We propose single chain type transient network models to explain
experimental results. Our models incorporate the spatial correlation
effect of micellar cores and average number of elastically active
chains per micellar core (the network functionality).
Our models can reproduce non-single Maxwellian relaxation and nonlinear
rheological behavior such as the shear thickening and thinning.
They are qualitatively consistent with experimental results.
In our models, the linear rheological behavior is mainly attributable to the difference of
network structures (functionalities).
The nonlinear rheological behavior is attributable to the
nonlinear flow rate dependence of the spatial correlation of micellar
core positions.
\end{abstract}
%
%
%

\maketitle


\section{Introduction}

Telechelic associative polymers which have hydrophilic main chains and
hydrophobic
chain ends form self-assembled micellar structures in aqueous
solutions\cite{Larson-book,Annable-Buscall-Ettelaie-Whittelstone-1993,Tripathi-Tam-McKinley-2006}.
If the concentration of telechelic polymers is sufficiently
high, the micellar cores are bridged by polymers to form a percolated,
network like structure. Such a network structure shows various kinds of interesting
rheological behavior and rheological properties of telechelic
associative polymer solutions have been extensively studied.
For example, various experimental results have been reported for
hydrophobically modified ethoxylated urethane (HEUR) aqueous solutions
\cite{Annable-Buscall-Ettelaie-Whittelstone-1993,Xu-Yekta-Li-Masoumi-Winnik-1996,Tam-Jenkins-Winnik-Bassett-1998,Tripathi-Tam-McKinley-2006}.

One characteristic rheological property of telechelic polymer solutions is
the linear viscoelasticity being well described by a single Maxwellian
form in a rather wide range of experimental parameters.
Theoretically, such a rheological
property can be explained in the framework of the transient network
model\cite{Green-Tobolsky-1946,Tanaka-Edwards-1992}.
The characteristic relaxation time is considered to be
the dissociation time of a chain end, and the characteristic modulus is
considered to be the elastic modulus of the network formed by telechelic polymers.
However, the simple description by the transient network model is not always valid. For relatively
high concentration solutions, deviation from the single Maxwellian form
has been reported\cite{Cathebras-Collet-Viguier-Berret-1998,Serero-Aznar-Porte-Berret-Calvet-Collet-Viguer-1998,Serero-Jacobsen-Berret-2000}.
In most of the transient network type models, such non-single Maxwellian
relaxation cannot be explained straightforwardly. (This is due to the
nature of the mean field type approximations in the
transient network type models.) This implies that we need to
incorporate some additional factors, which are affected by the polymer
concentration, to the models.

Another characteristic rheological property is that telechelic polymer
solutions often exhibit shear thickening behavior.
The transient network model has been improved to reproduce
such experimentally observed nonlinear rheological behavior.
In some improved versions of the transient network model\cite{Wang-1992a,Marrucci-Bhargava-Cooper-1933,Vaccaro-Marrucci-2000,Indei-Koga-Tanaka-2005,Indei-2007,Indei-2007a,Koga-Tanaka-2010,Tripathi-Tam-McKinley-2006}, the nonlinear
elasticity (or finite extensively) and the stretch dependent
dissociation rate are introduced. It has been reported that the
combination of the nonlinear elasticity and the stretch dependent
dissociation rate can reproduce several nonlinear rheological
properties.
For example, models by Tripathi {\it et al.}\cite{Tripathi-Tam-McKinley-2006} and
by Indei, Koga and Tanaka\cite{Indei-Koga-Tanaka-2005,Indei-2007,Indei-2007a,Koga-Tanaka-2010} can reproduce several
nonlinear features such as the shear thickening and thinning, by
tuning some fitting
parameters (which represent the strengths of the nonlinear elasticity and
the stretch dependent dissociation).

However, a recent study showed that conventionally proposed transient
network models cannot explain some
rheological data of a
HEUR aqueous solution\cite{Suzuki-Uneyama-Inoue-Watanabe-2012}, such
as the first normal stress coefficient ($\Psi_{1}$) data that exhibit no
nonlinearity in the shear-thickening regime of the shear viscosity ($\eta$).
To explain this experimental fact, a new transient network model with
an anisotropic (and shear rate dependent) bridge formation process was
proposed.
The anisotropic bridge formation model can
qualitatively explain this behavior of $\Psi_{1}$ as well as the shear thickening (and thinning)
behavior of $\eta$ under fast shear, the latter being
explained also by conventionally proposed models.
Although the new model provides a new
viewpoint to understand dynamical behavior of telechelic polymer solutions,
it was constructed in a phenomenological way and the molecular level
mechanism of anisotropic bridge formation has not been specified explicitly.

From the viewpoint of nonequilibrium dynamics of
flowing systems, anisotropies under flow are physically reasonable. For
example, colloid dispersions show structural anisotropies and
nonlinear rheological behavior under flow\cite{Verduin-deGans-Dhont-1996,Brader-2010}. The interaction
potential and spatial correlation between colloid particles are
important to understand/analyze such systems.
In telechelic polymer solutions, there are also interactions
and spatial correlations between micellar
cores\cite{Semenov-Joanny-Khokhlov-1995,Khalatur-Khokhlov-1996,Serero-Aznar-Porte-Berret-Calvet-Collet-Viguer-1998,Serero-Jacobsen-Berret-2000}
(although their forms are quite different from colloid dispersions).
Clearly, such interactions or spatial correlations depend on the
polymer concentration, and thus we expect the rheological properties of
telechelic polymer solutions to be rather strongly dependent on the
concentration. Indeed, we can find rather strong concentration
dependence of several rheological properties in literature\cite{Annable-Buscall-Ettelaie-Whittelstone-1993,Xu-Yekta-Li-Masoumi-Winnik-1996}.

This implies that we should design a model for telechelic polymer
solutions which directly takes into account the spatial correlation (or the
interaction potential) of the micellar cores, to
naturally reproduce concentration dependence of rheological properties
and anisotropic correlations and nonlinear rheological
behavior under shear. Nevertheless, the importance of the spatial
correlations for the rheology of telechelic polymer solutions has been
rarely pointed out.
Especially, the spatial correlations have been overlooked or ignored for
single chain type models, as far as the authors know.

In this work, we conducted linear viscoelastic measurements for
HEUR aqueous solutions with different concentrations. We show that the
linear rheological behavior is strongly affected by the concentration. The experimental
data suggest that there are two different concentration regimes.
Then we propose a simple but non-trivial single chain type
transient network models for telechelic associative polymers, which
explicitly take into account the spatial correlation information.
We show that our models give both single
and non-single Maxwellian relaxation moduli depending on
the polymer concentration.
We also show that nonlinear
rheological behavior such as the shear thickening and thinning can
be reproduced by our models considering neither nonlinear elasticity nor stretch dependence of the
bridge reconstruction rates.
We discuss concentration dependence of rheological data from the view
point of the transient network type models.

\section{Experimental}
\label{experimental}

\subsection{Materials and Measurements}

HEUR with hexadecyl end groups and polyethylene oxide (PEO) main chain was
used. The HEUR was synthesized
previously\cite{Suzuki-Uneyama-Inoue-Watanabe-2012} with a conventional
method, and the molecular
weight and polydispersity index were $M_{w} = 4.6
\times 10^{4}$ and $M_{w} / M_{n} = 1.35$
(the PEO precursor molecular weight and polydispersity index were $M_{w} = 1.9
\times 10^{4}$ and $M_{w} / M_{n} = 1.1$). The details of the synthesis
and characteristics are found in Ref.~\cite{Suzuki-Uneyama-Inoue-Watanabe-2012}.

The HEUR aqueous solutions were prepared by dissolving HEUR into
distilled water through stirring for $24 \text{ hours}$. For comparison, the
PEO aqueous solutions with a similar molecular weight and the same
concentrations as the HEUR solutions were also prepared. The PEO sample was purchased from Aldrich ($M_{w}
= 3.9 \times 10^{4}$, $M_{w} / M_{n}=1.08$) and dissolved into distilled
water.

Storage and loss moduli, $G'(\omega)$ and $G''(\omega)$, were measured for HEUR aqueous solutions at
various temperatures and concentrations. The measurements were conducted
with a stress-controlled rheometer (MCR-301, Anton Paar). A cone-plate
type fixture with the diameter $d = 75 \text{ mm}$ and the cone angle
$\theta = 1^{\circ}$ was utilized.
The measurement temperatures were $T = 5, 10, 15, 20,$ and
$25^{\circ}\text{C}$, and the concentration $c$ was
varied from $1\text{wt\%}$ to $10\text{wt\%}$. The angular frequency
range was set as $0.03$ rad/s $\le \omega \le$ $500$ rad/s.
Storage and loss moduli of PEO aqueous solutions with
the same concentrations and the similar molecular weight as HEUR solutions were also
measured in the same condition. (Due to the low moduli of solutions and the limitation of the equipment, only the loss moduli
data in a limited range $1 \text{ rad/s} \lesssim \omega \lesssim 10
\text{ rad/s}$ were obtained with acceptable accuracy.)

\subsection{Results}

First we show the master curves
for the storage and loss moduli,
$G'(\omega)$ and $G''(\omega)$ of $1\text{wt\%}$ HEUR aqueous solution
at the reference temperature $T_{r} = 25^{\circ}\text{C}$
in Figure \ref{heur_1wt_master_curves_slow}. The master curves
(master plots of $b_{T,0}^{-1} G'$ and $b_{T,0}^{-1} G''$ against $ a_{T,0} \omega$) are
constructed by shifting $G'(\omega)$ and $G''(\omega)$ data at
different temperatures horizontally and vertically, to superimpose the low
frequency data. The horizontal and vertical shirt factors
are denoted by $a_{T,0}$ and $b_{T,0}$, respectively.
As reported in the previous work\cite{Suzuki-Uneyama-Inoue-Watanabe-2012}, the time-temperature
superposition works well for this HEUR solution and the $G'(\omega)$ and
$G''(\omega)$ data are fitted well by a single Maxwell model, except the
high frequency region.
\begin{align}
 G'(\omega) = G_{0} \frac{(\tau \omega)^{2}}{1 + (\tau \omega)^{2}}, \qquad
 G''(\omega) = G_{0} \frac{\tau \omega}{1 + (\tau \omega)^{2}}
\end{align}
Here $\tau$ and $G_{0}$ are the characteristic relaxation time and
modulus, respectively.
The deviation in the high frequency region means
that there is another relaxation mode of which temperature dependence is
different from the relaxation mode in the low frequency region. In the
followings, we call
these modes in the high and low frequency regions as fast and slow modes, respectively.

\begin{figure}[h]
 \centering
 {\includegraphics[width=0.95\linewidth,clip]{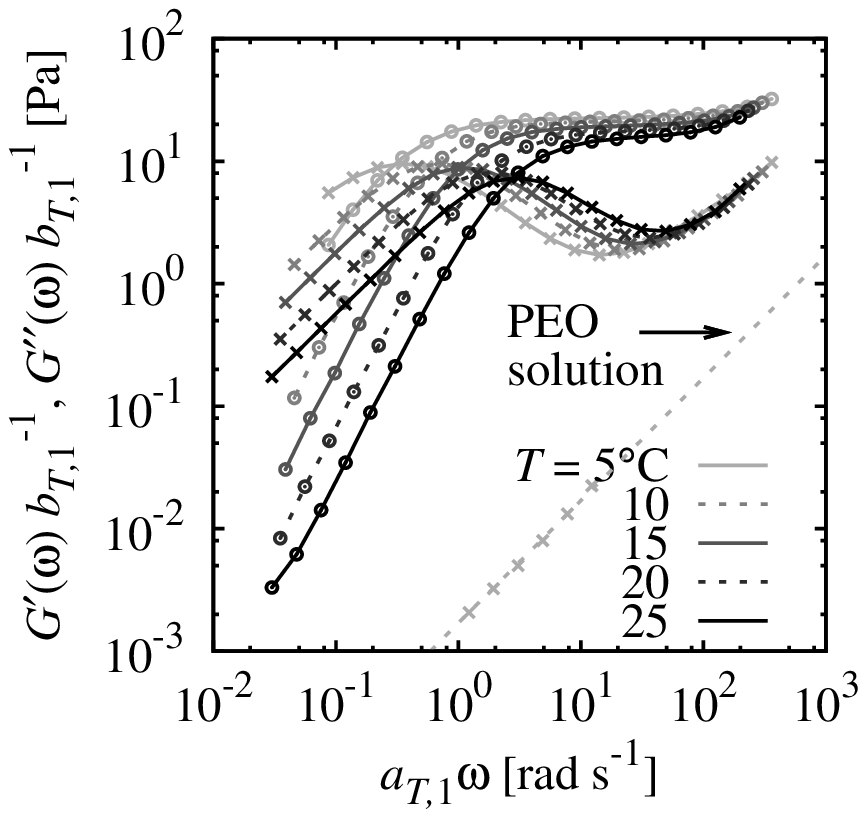}}
 \caption{\label{heur_1wt_master_curves_fast}}
\end{figure}

\begin{figure}[h]
 \centering
 {\includegraphics[width=0.95\linewidth,clip]{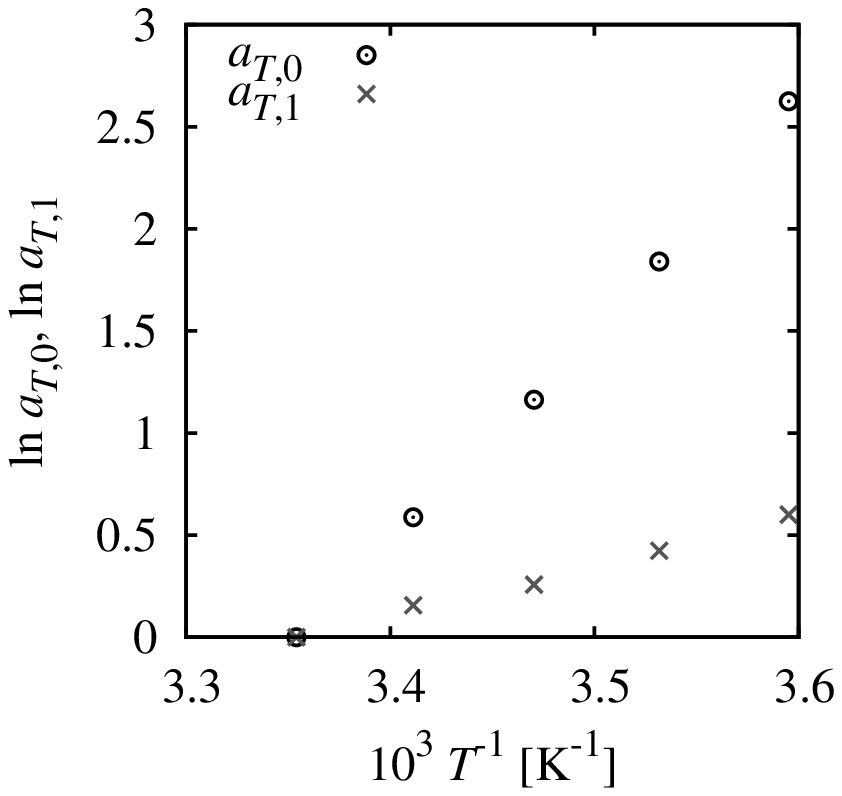}}
 \caption{\label{heur_1wt_shift_factor}}
\end{figure}

Ng {\it et al.}\cite{Ng-Tam-Jenkins-2000}
proposed that the fast mode is due to the relaxation of micellar
cores, whereas Bedrov {\it et al.}\cite{Bedrov-Smith-Douglas-2002} claimed that
the fast mode corresponds to the local
relaxation of hydrophilic main chains. If the fast mode corresponds to the
local relaxation mode of PEO chains, it should show the same temperature
dependence as a PEO aqueous solution without associative end groups.
We determined the horizontal shift factor $a_{T,1}$ of the
$1\text{wt\%}$ PEO solution (utilizing the vertical shift factor
$b_{T,1} = T / T_{r}$).
Then we shifted the $G'$ and $G''$ data of the $1\text{wt\%}$ HEUR solution with
the shift factors $a_{T,1}$ and $b_{T,1}$. The result is shown in
Figure \ref{heur_1wt_master_curves_fast}, and $a_{T,1}$ is compared with
$a_{T,0}$ of the HEUR solution in Figure \ref{heur_1wt_shift_factor}.
As shown in Figure
\ref{heur_1wt_master_curves_fast}, $a_{T,1}$ allows the
$G'(\omega)$ and $G''(\omega)$ data to
collapse into master curves at high frequencies where the fast mode is observed.
$a_{T,1}$ is less dependent on $T$ compared with $a_{T,0}$.
This result means that the fast mode corresponds to the local relaxation of PEO
chains, as claimed by Bedrov {\it et al.}\cite{Bedrov-Smith-Douglas-2002}.
However, it should be also noticed that
the relaxation time and the characteristic modulus of the fast mode of
the HEUR solution are
much larger than those of the PEO solution with the same
concentration, mostly because the HEUR chains form an associated
network-like structure and the relaxation modes of such a network are
different from ones of the free chains. We will discuss this point later.
Although we have shown the master curves only for the $1\text{wt\%}$
solution case, we can similarly construct master curves for slow and fast modes
for other concentration cases.

\begin{figure}[h]
 \centering
 {\includegraphics[width=0.95\linewidth,clip]{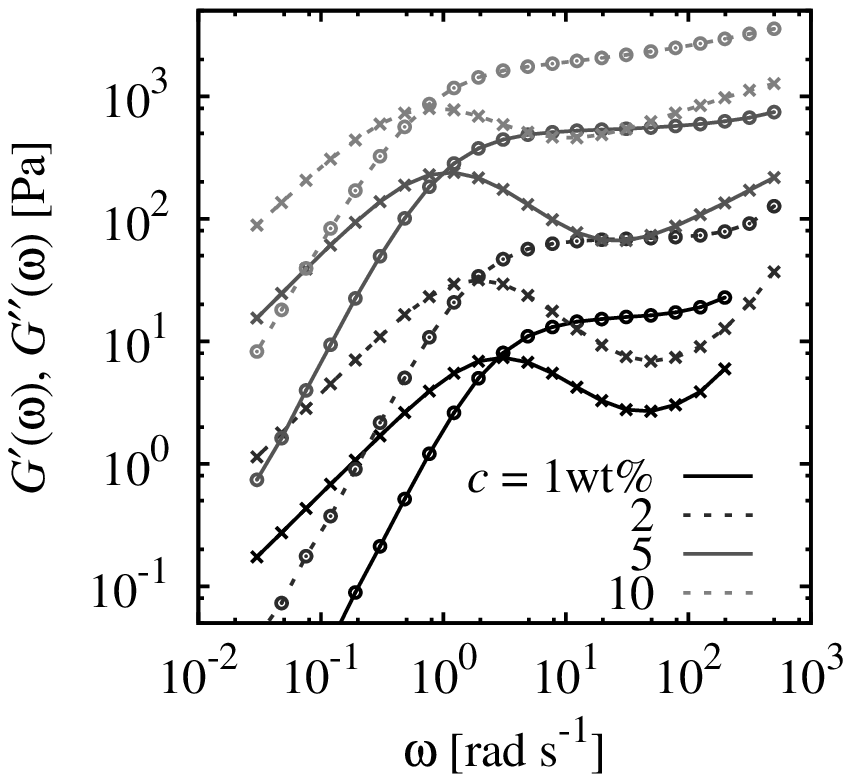}}
 \caption{\label{heur_concentration_dependence}}
\end{figure}

\begin{figure}[h]
 \centering
 {\includegraphics[width=0.95\linewidth,clip]{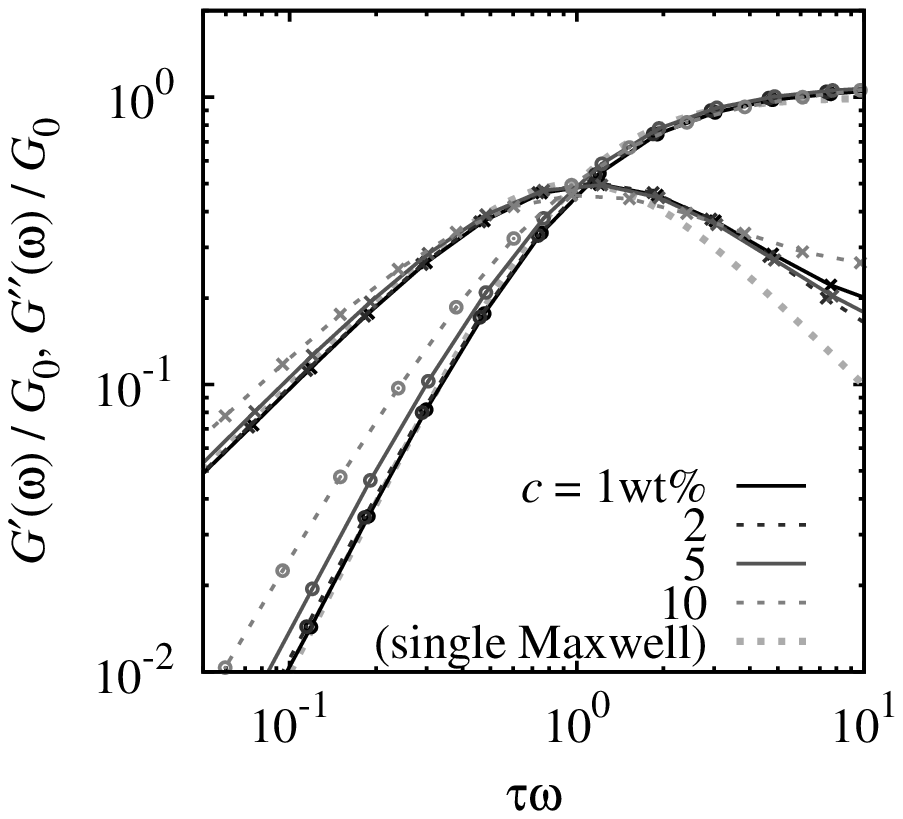}}
 \caption{\label{heur_concentration_dependence_rescaled}}
\end{figure}

We show the storage and loss moduli data at $T =
25^{\circ}\text{C}$ for different
concentrations ($c = 1, 2, 5,$ and $10\text{wt\%}$) in Figure
\ref{heur_concentration_dependence}. We can observe slow and fast modes
for all concentrations. However, the $\omega$ dependence of $G'(\omega)$ and
$G''(\omega)$ changes with the concentration rather largely.
This change is more prominent for the fast mode than for the slow mode.
For $c = 1$ and $2\text{wt\%}$, the $G''(\omega)$ at high $\omega$ (the
fast mode) is almost proportional to $\omega$, which indicates that the fast and slow modes are rather
separated. For $c = 5$ and $10\text{wt\%}$, on the other hand,
the $G''(\omega)$ at high $\omega$ becomes less dependent on $\omega$, and 
the fast and slow modes are not well separated compared with the
cases of the $1$ and $2\text{wt\%}$ solutions.
The slow mode distribution also changes with the concentration, although it
is not so clear in Figure \ref{heur_concentration_dependence}. To see
this change clearly, we show the $G'(\omega)$
and $G''(\omega)$ rescaled by the characteristic time $\tau$ and
modulus $G_{0}$ for the slow mode in
Figure \ref{heur_concentration_dependence_rescaled}. It is evident
from Figure \ref{heur_concentration_dependence_rescaled} that the slow
mode deviates from the single Maxwellian form as the concentration
increases. Similar non-single Maxwellian 
relaxation moduli have already been reported in some experimental or simulation
studies\cite{Cathebras-Collet-Viguier-Berret-1998,Serero-Aznar-Porte-Berret-Calvet-Collet-Viguer-1998,Serero-Jacobsen-Berret-2000,Tae-Kornfield-Hubbell-Lal-2002,Berret-Calvet-Collet-Viguier-2003,Castelleto-Hamley-Xue-Sommer-Pedersen-Olmsted-2004,Cass-Heyes-Blanchard-English-2008,Sprakel-Spruijt-vanderGucht-Padding-Briels-2009}.
For example, Berret and coworkers
\cite{Cathebras-Collet-Viguier-Berret-1998,Serero-Aznar-Porte-Berret-Calvet-Collet-Viguer-1998,Serero-Jacobsen-Berret-2000}
reported that the shear relaxation moduli of solutions of F-HEUR (HEUR with
perfluoroalkyl chain ends) can be fitted by the stretched exponential
form, $G(t) = G_{0} \exp[- (t / \tau)^{\alpha}]$ (with $G_{0}$, $\tau$ and
$\alpha$ being the characteristic modulus, characteristic relaxation
time, and the power-law exponent). Mistry {\it et al.} reported
non-single Maxwell relaxation modulus data for relatively high
concentration poly(butylene oxide)-(etylene oxide)-(butylene oxide) triblock copolymer solutions\cite{Mistry-Annable-Yuan-Booth-2006}.
Some shear relaxation modulus data
for relatively high concentration HEUR solutions\cite{Annable-Buscall-Ettelaie-Whittelstone-1993,Xu-Yekta-Li-Masoumi-Winnik-1996}
also deviate from the single Maxwellian form.

\begin{figure}[h]
 \centering
 {\includegraphics[width=0.8\linewidth,clip]{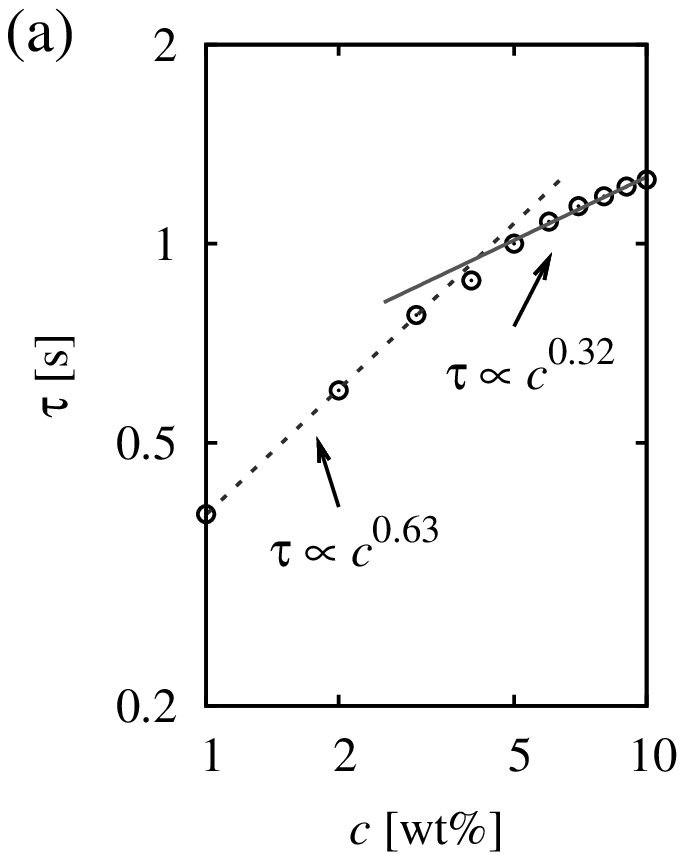}}
 {\includegraphics[width=0.8\linewidth,clip]{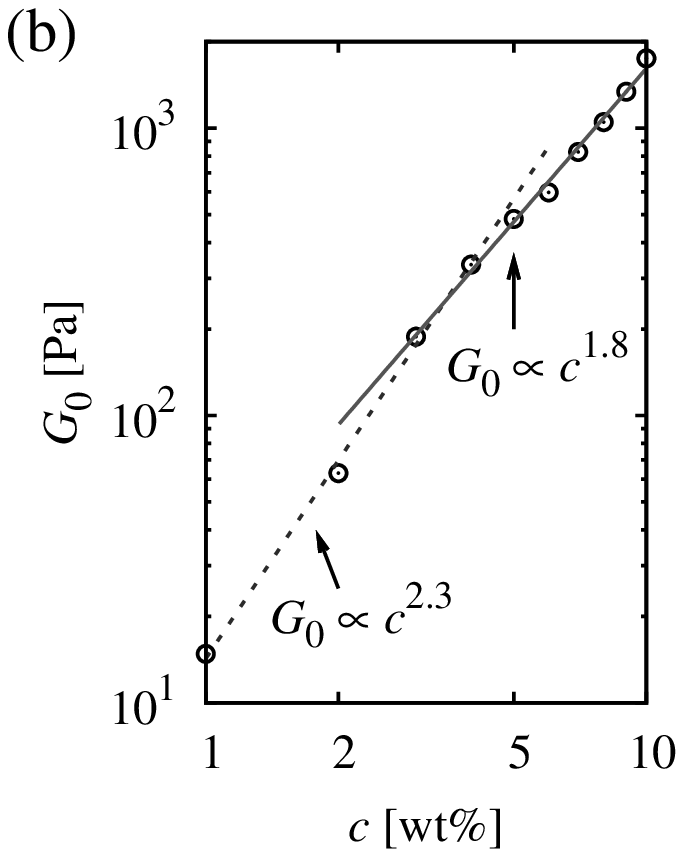}}
 \caption{\label{heur_tau0_g0_concentration_dependence}}
\end{figure}

Finally we show the characteristic relaxation time $\tau$ and
modulus $G_{0}$ data of the slow mode in Figure
\ref{heur_tau0_g0_concentration_dependence}. Both $\tau$ and $G_{0}$
depend on the HEUR concentration. The concentration dependence at low and
high concentration regions can be individually fitted to power laws. The
characteristic time $\tau$ depends on the concentration $c$ as
$\tau \propto c^{0.63}$ and $\tau \propto c^{0.32}$ for low and
high concentration regions, respectively (Figure \ref{heur_tau0_g0_concentration_dependence}(a)). Similarly, the characteristic
modulus depends on $c$ as $G_{0} \propto c^{2.3}$ and $G_{0} \propto
c^{1.8}$ (Figure \ref{heur_tau0_g0_concentration_dependence}(b)). Both $\tau$ and $G_{0}$ data show stronger $c$-dependence
at the low concentration region. From Figure
 \ref{heur_tau0_g0_concentration_dependence}, the crossover concentration $c_{c}$ is roughly
estimated as $c_{c} \approx 4\text{wt\%}$. At $c > c_{c}$,
the $G'(\omega)$ and $G''(\omega)$ data deviate
considerably from the single Maxwellian forms, as noted in Figure \ref{heur_concentration_dependence_rescaled}.
Here it should be noted that Annable {\it et al.}\cite{Annable-Buscall-Ettelaie-Whittelstone-1993} already measured the
concentration dependence of $\tau$ and $G_{0}$ of HEUR solutions
systematically, and reported similar data. (They proposed a statistical
model to explain their experimental data. Their model takes account of
superbridge and superloop structures, and can reasonably reproduce the
concentration dependence of $\tau$ and $G_{0}$.)
Further discussions of the experimental results in relation to
conventional transient network models will be made in Sec.~\ref{discussion}.

\section{Theoretical Model}
\label{theoretical_model}

As we have shown in Section \ref{experimental}, the linear
viscoelasticity of a telechelic polymer solution depends on the
concentration and there are two different concentration regimes.
The nonlinear rheological data of the $1\text{wt\%}$ aqueous solution of
the same HEUR has suggested the
existence of the anisotropic bridge formation dynamics under flow\cite{Suzuki-Uneyama-Inoue-Watanabe-2012}.
To explain such rheological
behavior by theoretical models, we need to employ models which can
take into account the polymer concentration and some spatial correlation
between micellar cores.
One possible candidate is the statistical model by Annable {\it et al.}\cite{Annable-Buscall-Ettelaie-Whittelstone-1993},
which can reproduce concentration dependence of linear rheological
properties. However, it is not easy to study nonlinear rheological
properties by their model. We consider that the transient network type
models are suitable to study both the linear and nonlinear rheological
properties.
In this section we propose
transient network type models which take into account effects of the average number of bridge
chains per micellar core and the correlation between micellar cores.
We show that these factors qualitatively affect several rheological
properties. We consider only the slow mode, and thus the models in the followings can
not reproduce the fast mode.

\subsection{Dense Network Model}
\label{dense_network_model}

\begin{figure}[h]
 \centering
 {\includegraphics[width=0.9\linewidth,clip]{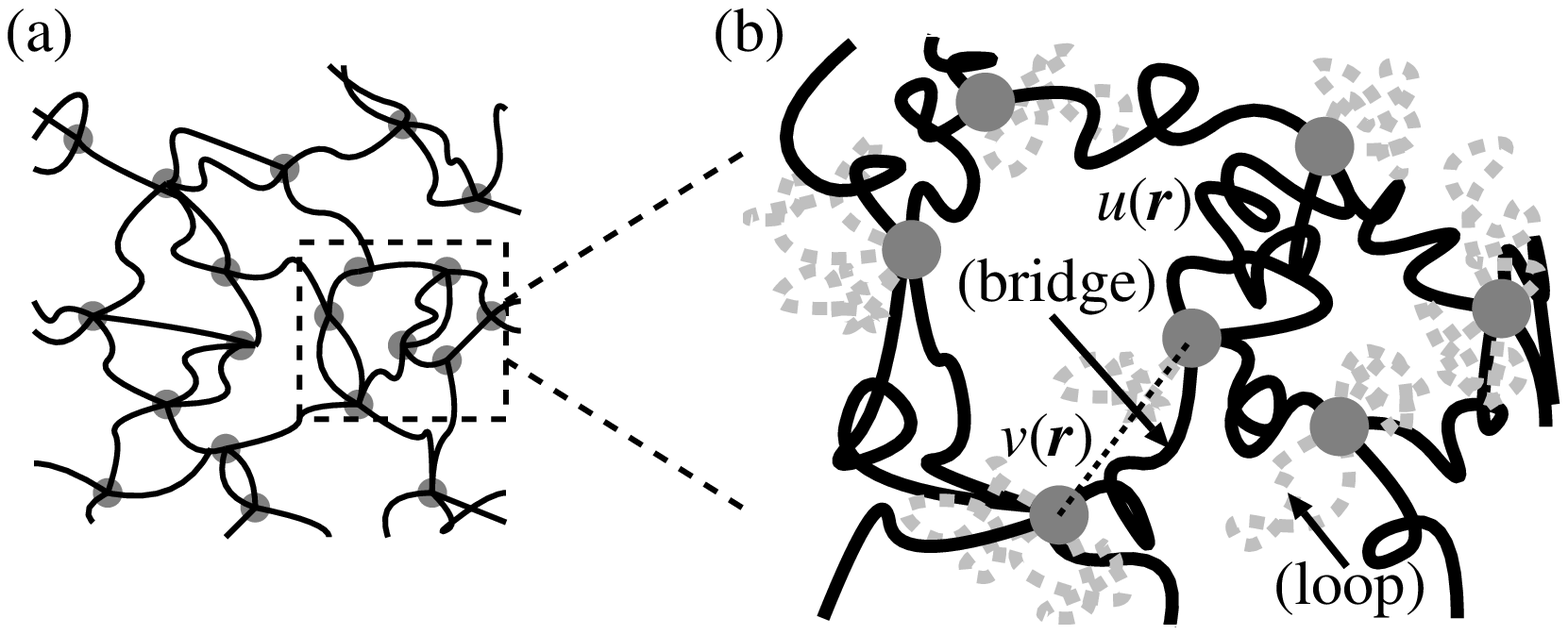}}
 \caption{\label{schematic_image_of_single_chain_model}}
\end{figure}

We first consider cases where the average number of bridge chains per micellar core
(the functionality $f$) is sufficiently large (typically $f \gtrsim 3$)
and dense networks are formed. A schematic image of a well-percolated network is
shown in Figure \ref{schematic_image_of_single_chain_model}(a).
We assume that the polymer chains are not sufficiently long and not too much concentrated
so that the entanglement effects are negligible.
For such cases, it will be sufficient to
consider only one tagged (labeled) chain in the system. In the
followings, we call such networks as ``the dense networks'' and the model
described in this subsection as ``the dense network model''.

Before we consider the dynamics, we focus on the equilibrium
probability distribution function for a single, tagged chain in the system.
We assume that a polymer chain can take two
states; the bridge and the loop. (The dangling chains are not explicitly considered here
because the characteristic reassociation time scale of the dangling
chains should be much shorter
than those of the bridge or loop states.) A schematic image is depicted
in  Figure \ref{schematic_image_of_single_chain_model}(b).
We express the state of the
chain by a state variable $n$, which takes $0$ (loop) or $1$ (bridge).
If the chain is in the bridge state, the information of the end-to-end
vector $\bm{r}$ is needed to fully specify the state. We can interpret
$\bm{r}$ as the relative position of the partner core, as shown in
Figure \ref{schematic_image_of_single_chain_model}(b).

Under the mean field type approximation, the interaction between two
micellar cores is expressed by an effective mean field interaction
potential $v(\bm{r})$. The effective interaction potential consists of
several different effects such as the steric repulsion (due to the corona
chains) and hard core
like interaction between micellar cores. (We expect that the steric repulsion will be dominant, and practically other contributions may be ignored.) If a bridge is
formed, the chain feels an extra potential energy. Roughly speaking, this
extra energy corresponds to the
elastic free energy of a chain, $u(\bm{r})$, and
the equilibrium probability distribution is given as
\begin{align}
 & \label{equilibrium_probability_distribution_p0_single_chain}
  P_{\text{eq}}(0) = \frac{1}{\Xi} \mathcal{Z}_{0} \\
 & \label{equilibrium_probability_distribution_p1_single_chain}
  P_{\text{eq}}(1,\bm{r}) = \frac{\mathcal{Q}_{1}}{\Xi \Lambda^{3}} \exp
  \left[ - \frac{v(\bm{r}) + u(\bm{r})}{k_{B} T} \right]
\end{align}
Here $\Lambda$ is the thermal de Broglie wave length, and $\Xi$ is the
partition function. $\mathcal{Z}_{0}$ and $\mathcal{Q}_{1}$ are the partition function of a
single loop chain and the partial partition function of a single bridge chain,
respectively. The quantities with the subscript ``eq'' represent the equilibrium
quantities.
\begin{equation}
 \label{partition_function}
  \Xi
  \equiv \mathcal{Z}_{0} + \frac{\mathcal{Q}_{1}}{\Lambda^{3}} \int d\bm{r} \,
  \exp \left[ - \frac{v(\bm{r}) + u(\bm{r})}{k_{B} T} \right] \\
\end{equation}

The effective interaction potential can be related to the structure of
the multi chain system. It should be determined so that the resulting
spatial structure becomes consistent with the structure of a target
system, such as the structure detected by scattering experiments.
Under the mean field approximation, the radial distribution function
(RDF) of the system can be easily calculated.
\begin{equation}
\begin{split}
 \label{rdf_and_effective_potential}
 \rho_{0} g_{\text{eq}}(\bm{r})
 & =
  n_{0} P_{\text{eq}}(1,\bm{r}) + \rho_{0} e^{ - v(\bm{r}) / k_{B} T}
 P_{\text{eq}}(0) \\
 & = \frac{\rho_{0} \mathcal{Z}_{0}}{\Xi}
 [ 1 + \xi e^{- u(\bm{r}) / {k_{B} T}} ]
  e^{ - {v(\bm{r})} / {k_{B} T} }
\end{split}
\end{equation}
Here $\rho_{0}$ is the average spatial density of the micellar core,
$n_{0}$ is the aggregation number of a micellar core, and $\xi \equiv n_{0} \mathcal{Q}_{1} / \rho_{0} \Lambda^{3} \mathcal{Z}_{0}$
is the effective activity (fugacity). We have assumed that the system
size is sufficiently large.
Eq \eqref{rdf_and_effective_potential} gives the relation between the mean field potential and the radial
distribution function.
\begin{equation}
 \label{effective_potential_explicit}
 v(\bm{r}) 
  = - k_{B} T \ln g_{\text{eq}}(\bm{r})
  + k_{B} T \ln [ 1 + \xi e^{- {u(\bm{r})} / {k_{B} T}} ]
\end{equation}
Here we have dropped an additive constant which does not affect
thermodynamic properties.
The first term in the right hand side of
\eqref{effective_potential_explicit} is the effective potential form that
appears under some
approximations (for example, in the liquid state theory\cite{Hansen-McDonald-book}).
The second term in the right hand side of
\eqref{effective_potential_explicit} represents the repulsive
interaction between cores which cancels the attractive interaction by
bridges. We cannot reproduce the correct spatial correlation without
this repulsive interaction.

For convenience, we introduce another equilibrium probability
distribution function. We express the (unnormalized) conditional probability distribution of a
micellar core under the condition $n = 0$ as
\begin{equation}
 \label{equilibrium_probability_distribution_phi}
 \Phi_{\text{eq}}(\bm{r}) \equiv
  \frac{\rho_{0}}{n_{0}}
  e^{ - {v(\bm{r})} / {k_{B} T} }
\end{equation}
It is straightforward to show that the following relation holds.
\begin{equation}
 P_{\text{eq}}(1,\bm{r})
   = \xi e^{-u(\bm{r}) / k_{B} T} \Phi_{\text{eq}}(\bm{r}) P_{\text{eq}}(0)
\end{equation}
Eq \eqref{equilibrium_probability_distribution_phi} is useful when we
consider the bridge construction process.
This is because $\Phi_{\text{eq}}(\bm{r})$ represents the probability to find a partner core at a certain
position $\bm{r}$ in space, which is required when we consider the bridge
construction process.

To study dynamical properties such as rheological properties, we need
the dynamic equations for probability distributions.
We consider the time evolution equations for three time-dependent
distribution functions ($P(1,\bm{r},t)$, $P(0,t)$, and $\Phi(\bm{r},t)$).
Roughly speaking, there are two different contributions for dynamic
equations. One is the reconstruction of bridge chains, and another is
the motion of micellar cores.

First we consider the bridge reconstruction dynamics.
So far, several different destruction rate models have been proposed for
transient network type
models\cite{Tanaka-Edwards-1992,Vaccaro-Marrucci-2000,Indei-Koga-Tanaka-2005,Tripathi-Tam-McKinley-2006}.
Here we describe the destruction rate in the following form.
\begin{equation}
 \label{construction_rate_single_chain}
 W(0 | 1,\bm{r}) = \frac{1}{\tau(\bm{r})}
\end{equation}
$\tau(\bm{r})$ represents the characteristic destruction time for the
bridge chain with the end-to-end vector $\bm{r}$. (In the next section,
we simply set $\tau(\bm{r})$ as a constant.)
The bridge formation rate should be proportional to the probability that
we find a partner core. Then the bridge formation rate can be
modeled as $W(1,\bm{r} | 0) \Phi(\bm{r},t)$.
From the detailed balance condition, the explicit form of $W(1,\bm{r} | 0)$ is
automatically determined.
\begin{equation}
  \label{destruction_rate_single_chain}
 W(1,\bm{r} | 0) = \frac{W(0 | 1,\bm{r}) P_{\text{eq}}(1,\bm{r})}
  {\Phi_{\text{eq}}(\bm{r}) P_{\text{eq}}(0)} = \frac{1}{\tau(\bm{r})}
  \xi e^{-u(\bm{r}) / k_{B} T}
\end{equation}
In most of conventional transient network models, the
dangling-to-bridge transition rate is utilized instead of the
bridge construction (loop-to-bridge transition) rate.
This difference is not serious, because
the dangling-to-bridge transition rate can be cast into the bridge
construction rate, unless the system is subjected to a very fast flow.

Here it is worth noting that in many transient network models,
both the dangling-to-bridge and
bridge-to-dangling transition rate models have been proposed and utilized to
reproduce rheological data well.
In other words, both two transition rates are freely tunable in these models.
However, as long as the
equilibrium probability distribution is given, the bridge construction
rate is automatically determined and no longer freely tunable.
(If the bridge construction rate is given, the bridge destruction
rate is automatically determined from the detailed balance condition.)
Our bridge reconstruction rates shown above do not affect the
equilibrium probability distribution.

Next we consider the contribution of the motion of micellar cores.
Micellar cores move by the thermal noise (the Brownian motion) and by
externally imposed flows.
These effects can be modeled by the (generalized) Fokker-Planck operator.
The Fokker-Planck operator can depend on
whether the chain is in the bridge state or in the loop state, and thus
we employ an $n$-dependent Fokker-Planck operator, $\mathcal{L}(n,t)$ ($n = 0$
and $1$, for loop and bridge).
In or near equilibrium, under the Markov
approximation, the Fokker-Planck operator $\mathcal{L}(n,t)$ can be expressed simply as
\begin{equation}
 \label{fokker_planck_operator_equilibrium_single_chain}
 \mathcal{L}(n,t) P(\bm{r})
  \equiv \frac{1}{\zeta_{0}} \frac{\partial}{\partial \bm{r}} \cdot
  \left[ \frac{\partial [n u(\bm{r}) + v(\bm{r})]}{\partial \bm{r}} P(\bm{r})
   + k_{B} T \frac{\partial P(\bm{r})}{\partial \bm{r}} \right]
  - \frac{\partial}{\partial \bm{r}} \cdot
  \left[ \bm{\kappa}(t) \cdot \bm{r} P(\bm{r}) \right]
\end{equation}
where $\zeta_{0}$ is the effective friction coefficient and
$\bm{\kappa}(t)$ is the velocity gradient tensor.
The effective friction coefficient $\zeta_{0}$ can be related to the long
time self diffusion coefficient of a micellar core $D_{0}$ as
\begin{equation}
 D_{0} = \frac{2 k_{B} T}{\zeta_{0}}
\end{equation}
The numerical factor $2$ comes from a fact that $\bm{r}$ is the distance
between two micellar cores and they move almost independently in the
long time limit.

It should be noticed that eq
\eqref{fokker_planck_operator_equilibrium_single_chain} is not always valid.
For example, if the memory effect\cite{vanKampen-book} is not negligible, we need to
introduce the memory kernel.
In nonequilibrium states, such as under shear
flow, the dynamics will be drastically affected by external driving
forces due to flow. If the system largely deviates from equilibrium,
the effective potential can be different from the equilibrium
potential. Also, the mobility tensor (friction coefficient) can become
qualitatively different from the equilibrium form
\cite{McPhie-Daivis-Snook-Ennis-Evans-2001,Uneyama-Horio-Watanabe-2011}. Therefore, for nonequilibrium
cases, we may need to employ a nonequilibrium effective free energy and
an anisotropic mobility tensor. It is not a simple task
to model nonequilibrium dynamics, and in the current work we mainly considers
dynamics near equilibrium.
Fortunately, the explicit form of
$\mathcal{L}(n,t)$ does not severely affect the results of the following
analyses (at least qualitatively), and we do not further discuss these
effects here.

Finally, by using the bridge reconstruction rates and the Fokker-Planck
operators, we have the following dynamic equations for probability
distributions.
\begin{align}
 & \label{dynamic_equation_p1_single_chain}
 \frac{\partial P(1,\bm{r},t)}{\partial t}
  = \mathcal{L}(1,t) P(1,\bm{r},t)
 + W(1,\bm{r} | 0) \Phi(\bm{r},t) P(0,t) - W(0 | 1,\bm{r}) P(1,\bm{r},t) \\
 & \label{dynamic_equation_p0_single_chain}
 \frac{\partial P(0,t)}{\partial t}
  = \int d\bm{r} \, [W(0 | 1,\bm{r}) P(1,\bm{r},t) - W(1,\bm{r} | 0)
 \Phi(\bm{r},t) P(0,t)] \\
 & \label{dynamic_equation_phi_single_chain}
 \frac{\partial \Phi(\bm{r},t)}{\partial t}
  = \mathcal{L}(0,t) \Phi(\bm{r},t)
\end{align}
The dynamic equation for the probability distribution of the micellar
core, $\Phi(\bm{r},t)$ (eq
\eqref{dynamic_equation_phi_single_chain}), was not introduced in
the previous transient network
models\cite{Wang-1992a,Marrucci-Bhargava-Cooper-1933,Vaccaro-Marrucci-2000,Indei-Koga-Tanaka-2005,Indei-2007,Indei-2007a,Koga-Tanaka-2010,Tripathi-Tam-McKinley-2006}.
(The previous models implicitly assumed that $\Phi(\bm{r},t)$ does not depend on flow history.)
As we will show in the next section, $\Phi(\bm{r},t)$
largely affects rheological properties.

\subsection{Sparse Network Model}
\label{sparce_network_model}

\begin{figure}[h]
 \centering
 {\includegraphics[width=0.9\linewidth,clip]{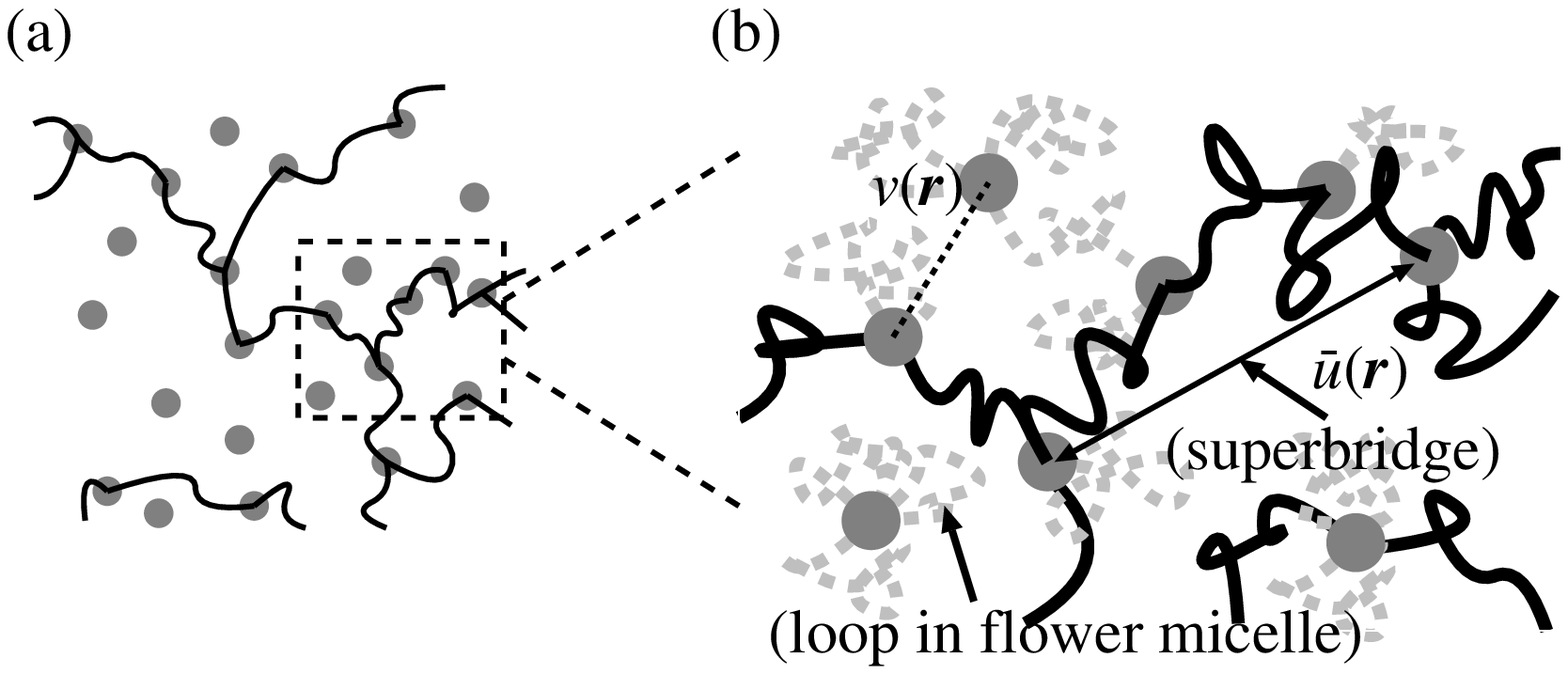}}
 \caption{\label{schematic_image_of_single_effective_bond_model}}
\end{figure}

If the functionality is relatively small ($f \lesssim 3$), the dense
network model introduced in the previous subsection is
not appropriate. In such cases, many (though not all) micellar cores in a network
link only two bridge chains and can not be regarded as active nodes
sustaining the network elasticity.
Then networks are mainly formed by so-called superbridge
structures\cite{Annable-Buscall-Ettelaie-Whittelstone-1993}, and the
(apparent) fraction of elastically active chains become small.
We call such networks as ``the sparse networks'' and the model
described in this subsection as ``the sparse network model''.
A schematic image is depicted
in Figure \ref{schematic_image_of_single_effective_bond_model}(a).

In the sparse network model, most of chains in the system is not
elastically independent
(See Figure \ref{schematic_image_of_single_effective_bond_model}(b)), and thus
we should use a
superbridge as the elementary unit (effective bond) to construct a mean
field model.
For simplicity, we assume that the number of bridge chains which form a
single superbridge is constant and express it as $m$ ($m \gtrsim 2$).
We assume the effective interaction potentials are the same
as ones in the dense network model. Even with these assumptions, it is
still difficult to accurately
formulate a mean field model for a single superbridge.
In the current work, we attempt to construct a model which captures
the essential properties, rather than a quantitatively accurate
model. As we will show later, many features of the sparse
network model are consistent with experiments even if we employ rough
approximations and simplifications.

We express the probability that a superbridge (with the end-to-end vector
$\bm{r}$) is present as $\bar{P}_{\text{eq}}(1,\bm{r})$.
As a rough estimate, this
can be approximately expressed in terms of the equilibrium
probability distribution of a bridge chain (in the dense network model, eq
\eqref{equilibrium_probability_distribution_p1_single_chain}).
In the followings, we utilize the short-hand notations for
convolutions.
\begin{align}
 & [f * g](\bm{r}) \equiv \int d\bm{r}' \, f(\bm{r} - \bm{r}')
 g(\bm{r}') \\
 & f^{[m]}(\bm{r}) \equiv \int d\bm{r}_{0} \dots d\bm{r}_{m}
 \, \delta(\bm{r} - \bm{r}_{m}) \delta(\bm{r}_{0}) \prod_{k = 1}^{m}
 f(\bm{r}_{k} - \bm{r}_{k - 1})
\end{align}
We approximately express $\bar{P}_{\text{eq}}(1,\bm{r})$ as
\begin{equation}
 \label{equilibrium_probability_distribution_p1_single_bond}
  \bar{P}_{\text{eq}}(1,\bm{r})
  \propto P_{\text{eq}}^{[m]}(1,\bm{r})
  \propto \frac{1}{\Lambda^{3 m}} \left[ e^{- (u + v) /
  k_{B} T} \right]^{[m]}(\bm{r})
\end{equation}

To make the expressions similar to the
dense network model, we introduce the following effective bond potential
$\bar{u}(\bm{r})$.
\begin{equation}
  \label{u_bar_definition}
 \bar{u}(\bm{r}) \equiv - v(\bm{r}) - k_{B} T \ln \left[ \frac{1}{\Lambda^{3 (m - 1)}} \left[ e^{- (u + v) /
  k_{B} T} \right]^{[m]}(\bm{r}) \right]
\end{equation}
Substituting eq \eqref{u_bar_definition}
into eq \eqref{equilibrium_probability_distribution_p1_single_bond}, we
can express
the equilibrium probability distributions in the
forms similar to eqs
\eqref{equilibrium_probability_distribution_p0_single_chain} and
\eqref{equilibrium_probability_distribution_p1_single_chain}.
\begin{align}
 & \label{equilibrium_probability_distribution_p0_single_bond}
 \bar{P}_{\text{eq}}(0) \approx
 \frac{\bar{\mathcal{Z}}_{0}}{\bar{\Xi}} \\
 & \label{equilibrium_probability_distribution_p1_single_bond_modified}
 \bar{P}_{\text{eq}}(1,\bm{r})
  \approx \frac{\bar{\mathcal{Q}}_{1}}{\Lambda^{3} \bar{\Xi}} \exp
 \left[ - \frac{\bar{u}(\bm{r}) + v(\bm{r})}{k_{B} T} \right]
\end{align}
where $\bar{\mathcal{Z}}_{0}$ is the partition function of an inactive
bond (a loop, a dangling end, or a superloop), and
$\bar{\mathcal{Q}}_{1}$ is the partial partition
function of an active superbridge. $\bar{\Xi}$ is the partition
function and defined as follows.
\begin{equation}
 \bar{\Xi} \equiv \bar{\mathcal{Z}}_{0} + \frac{\bar{\mathcal{Q}}_{1}}{\Lambda^{3}}
  \int d\bm{r} \, \exp
 \left[ - \frac{\bar{u}(\bm{r}) + v(\bm{r})}{k_{B} T} \right]
\end{equation}

Unlike the dense network model, the superbridge becomes elastically
inactive when one bridge chain in a superbridge is destructed.
The destruction rate is expected to be hardly dependent on the chain
stretch (unless the flow is very fast),
and thus we assume that the characteristic destruction time is constant.
(According to the Kramers theory\cite{vanKampen-book}, the destruction
rate is determined by the curvatures of the potential at the local
minima and maxima, and the energy barrier. Both of them are not largely affected
by the chain stretch.)
Thus the destruction rate of the superbridge becomes
\begin{equation}
 \label{destruction_rate_single_bond}
 \bar{W}(0 | 1,\bm{r}) \approx \frac{m}{\tau_{0}}
\end{equation}
with $\tau_{0}$ being the characteristic (or average) destruction time
of a single bridge chain. The superbridge construction rate is not so simple
compared with the dense network model.
Because the destruction event of an superbridge occurs when
one of the bridge chains in the superbridge is destructed, the construction
occurs when $(m - 1)$ bridge chains are present and a new bridge chain is
constructed. Such a consideration suggests the construction rate
in the following form.
\begin{equation}
 \label{construction_rate_single_bond}
   \bar{W}'(1, \bm{r} | 0) = \frac{1}{\Lambda^{3 (m - 1)}}
   \sum_{k = 1}^{m} \left[ (e^{-u / k_{B} T} \Phi)^{[m - k]} * 
  \Phi * (e^{-u / k_{B} T} \Phi)^{[k - 1]} \right](\bm{r})
\end{equation}
$\bar{W}'(1,\bm{r} | 0)$ is determined from the
detailed balance condition as
\begin{equation}
 \label{construction_rate_single_bond_explicit}
 \bar{W}'(1, \bm{r} | 0) 
  = \frac{m}{\tau_{0}} \bar{\xi}
  e^{- \bar{u}(\bm{r}) / k_{B} T}
  e^{[\bar{v}(\bm{r}) - v(\bm{r})] / k_{B} T}
\end{equation}
with
\begin{align}
 & \label{v_bar_definition}
 \bar{v}(\bm{r}) \equiv - k_{B} T \ln \bigg[ \frac{1}{\Lambda^{3 (m -
 1)}} \sum_{k = 1}^{m} [(e^{-(u + v) / k_{B} T})^{[m - k]} * 
  e^{-v / k_{B} T} * (e^{-(u + v) / k_{B} T})^{[k -
  1]}](\bm{r}) \bigg] \\
 & \label{xi_bar_definition}
 \bar{\xi} \equiv \frac{\bar{\mathcal{Z}}_{0}
  n_{0}^{m}}{\bar{\mathcal{Q}}_{1} \rho_{0}^{m} \Lambda^{3}}
\end{align}
In the followings, we express the summation of convolutions in eq
\eqref{construction_rate_single_bond} shortly as follows.
\begin{equation}
 \label{tilde_convolutions_definition}
 \widetilde{\Phi^{[m]}}(\bm{r}) \equiv \frac{e^{[\bar{v}(\bm{r}) - v(\bm{r})] /
 k_{B} T}}{\Lambda^{3 (m - 1)}}
  \sum_{k = 1}^{m} [(e^{-u / k_{B} T} \Phi)^{[m - k]} * 
  \Phi * (e^{-u / k_{B} T} \Phi)^{[k - 1]}](\bm{r})
\end{equation}
Also, it is convenient to introduce the following construction rate.
\begin{equation}
 \label{construction_rate_single_bond_modified}
 \bar{W}(1, \bm{r} | 0) 
 \equiv \bar{W}'(1, \bm{r} | 0)  e^{- [\bar{v}(\bm{r}) - v(\bm{r})] /
 k_{B} T} = \frac{m}{\tau_{0}} \bar{\xi}
  e^{- \bar{u}(\bm{r}) / k_{B} T}
\end{equation}
The construction rate is then rewritten as $\bar{W}(1,\bm{r} | 0)
\widetilde{\Phi^{[m]}}(\bm{r})$, which is formally similar to that in the dense
network model.

Under the Markov approximation, the Fokker-Planck operator for a
superbridge will be expressed as follows.
\begin{equation}
 \label{fokker_planck_operator_equilibrium_single_bond}
 \bar{\mathcal{L}}(1,t) P(\bm{r}) \approx
  \frac{1}{\bar{\zeta_{0}}} \frac{\partial}{\partial \bm{r}} \cdot
  \left[ \frac{\partial [\bar{u}(\bm{r}) + v(\bm{r})]}{\partial \bm{r}}
   P(\bm{r}) + k_{B} T \frac{\partial P(\bm{r})}{\partial \bm{r}}
  \right]
  - \frac{\partial}{\partial \bm{r}} \cdot
  \left[ \bm{\kappa}(t) \cdot \bm{r} P(\bm{r}) \right]
\end{equation}
Here $\bar{\zeta}_{0}$ is the effective friction coefficient for the
end-to-end vector of a superbridge. This effective friction coefficient
is related to the diffusion coefficient of elastically active nodes (to
which three or more active superbridges are attached). We expect that
$\bar{\zeta}_{0}$ is larger than $\zeta_{0}$, and thus the effect of the
Brownian motion will be small compared with the case of the dense network
model. Also, from eqs \eqref{destruction_rate_single_bond} and
\eqref{construction_rate_single_bond_modified}, the characteristic time
scale of the superbridge reconstruction process is relatively small. Therefore,
we expect that the effect of the Brownian motion will be negligibly
small for practical cases.

The dynamic equations for the sparse network model are finally
described as
\begin{align}
 & \label{dynamic_equation_p1_single_bond}
  \frac{\partial \bar{P}(1,\bm{r},t)}{\partial t}
  = \bar{\mathcal{L}}(1,t) \bar{P}(1,\bm{r},t)
  + \bar{W}(1, \bm{r} | 0) \widetilde{\Phi^{[m]}}(\bm{r},t) 
  \bar{P}(0,t)   - \bar{W}(0 | 1,\bm{r})
   \bar{P}(1,\bm{r},t) \\
 & \label{dynamic_equation_p0_single_bond}
  \frac{\partial \bar{P}(0,t)}{\partial t}
  = \int d\bm{r} \, \bigg[
  \bar{W}(0 | 1,\bm{r})
  \bar{P}(1,\bm{r},t) - \bar{W}(1, \bm{r} | 0) \widetilde{\Phi^{[m]}}(\bm{r},t) 
  \bar{P}(0,t) \bigg]
\end{align}
The dynamic equation of $\Phi(\bm{r},t)$ is the same as the dense network
model, eq \eqref{dynamic_equation_phi_single_chain}. While eqs \eqref{dynamic_equation_p1_single_bond} and
\eqref{dynamic_equation_p0_single_bond} are similar to eqs
\eqref{dynamic_equation_p1_single_chain} and
\eqref{dynamic_equation_p0_single_chain}, the dependence of the
construction rate on $\Phi(\bm{r},t)$ is qualitatively
different: $\Phi(\bm{r},t)$ in the dense network model is
replaced by a convolution
$\widetilde{\Phi^{[m]}}(\bm{r},t)$.

%

Before we proceed to analyses of rheological properties, we shortly
comment on the relations among the dense and sparse network models, and
the anisotropic bridge formation model.
Although we have assumed that a superbridge in the sparse network
model consists of multiple
bridge chains ($m \gtrsim 2$), it is formally possible to set $m =
1$. In the case of $m = 1$, convolutions
simply become $f^{[m]}(\bm{r}) = f(\bm{r})$ and
$\widetilde{\Phi^{[m]}}(\bm{r}) = \Phi(\bm{r})$. For this case, the sparse
network model reduces to the dense network model with $\tau(\bm{r})
= \tau_{0}$.

However, for $m > 1$, the sparse network model is
qualitatively different from the dense network model.
We need to employ the sparse network model if the
functionality is small and the contribution of superbridges is not negligible.
We expect that the dependence of the superbridge construction rate on
the spatial correlation is enhanced in the
sparse network model. This would be intuitively natural because
in the sparse network model, one superbridge consists of multiple bridge
chains connected in series and the superbridge conformation
strongly reflects the anisotropy of the underlying micellar core distribution.

The sparse network model reduces to the previously proposed anisotropic bridge
formation model, under some conditions. (The details are shown in
Appendix \ref{anisotropic_bridge_formation_model}.)
If we compare dynamic equations
\eqref{dynamic_equation_p1_single_bond},
\eqref{dynamic_equation_p0_single_bond} and
\eqref{dynamic_equation_phi_single_chain}
with the anisotropic bridge formation
model\cite{Suzuki-Uneyama-Inoue-Watanabe-2012}, we
find that $\Phi(\bm{r},t) P(0,t)$ works as the source function for
a newly constructed bridge.
The source function in the anisotropic bridge formation model was
introduced rather phenomenologically and its molecular meaning was left
rather arbitrary. In our present work, the source function is directly
related to the spatial distribution of the micellar cores.
Therefore, models formulated here lend support from a molecular point of
view, to some extent, to the use of the
phenomenologically designed model.

\subsection{Rheological Properties}

In this subsection we show some rheological properties calculated from our
models shown in Sections \ref{dense_network_model} and
\ref{sparce_network_model}. To make the models simpler and tractable, in
the followings we limit ourselves to simple systems.
We assume that effects of the nonlinear elasticity and the stretch dependent bridge
reconstruction are absent.
Although these effects are considered to be essential in some
transient network models, they are
not always necessary to reproduce experimentally observed
rheological properties. (We will discuss the effects of 
nonlinear elasticity and the stretch dependent bridge
reconstruction rates later.)

We employ the constant bridge destruction time model
\begin{equation}
 \tau(\bm{r}) = \tau_{0}
\end{equation}
where $\tau_{0}$ is a constant independent of $\bm{r}$.
(This assumption was already introduced for the sparse network model.)
We also employ the harmonic elastic potential of a bridge
chain (the Gaussian chain model).
\begin{equation}
 \label{linear_elasticity_potential}
 u(\bm{r}) = \frac{3 k_{B} T}{2 R_{0}^{2}} \bm{r}^{2}
\end{equation}
where $R_{0}$ is a constant which corresponds to the average
end-to-end distance of a chain.

To calculate rheological properties, we need the microscopic expression
of the stress tensor. In this work we assume that the stress is mainly caused
by the bridge chains.
For the dense network model, we define the microscopic stress tensor as
\begin{equation}
 \label{stress_tensor_definition_single_chain}
 \hat{\bm{\sigma}}(n) \equiv
  \begin{cases}
   \displaystyle \frac{n_{0} \rho_{0}}{2}
  \left[ \frac{\partial u(\bm{r})}{\partial \bm{r}} \bm{r}
   - k_{B} T \bm{1} \right] & (n = 1) \\
   0 & (n = 0)
  \end{cases}
\end{equation}
Here, $n_{0} \rho_{0} / 2$ is the number density of polymer chains and $\bm{1}$ is
the unit tensor.
This expression is consistent with the stress-optical rule because we
have employed a harmonic potential for $u(\bm{r})$ (eq \eqref{linear_elasticity_potential}).
It should be noted that eq \eqref{stress_tensor_definition_single_chain}
is not a unique candidate. For example, 
we may define the stress tensor in a way that it becomes a conjugate variable
to the deformation (In
this case, we have an extra term which is proportional to $[\partial
v(\bm{r}) / \partial \bm{r}] \bm{r}$. However, this term is not
important unless the micellar cores are rather concentrated.)
For a given set of probability distribution functions $P(1,\bm{r})$,
$P(0)$, and $\Phi(\bm{r})$, the average stress tensor is calculated as
(cf. eqs \eqref{linear_elasticity_potential} and \eqref{stress_tensor_definition_single_chain})
\begin{equation}
  \bm{\sigma}
  = \int d\bm{r} \, \hat{\bm{\sigma}}(1) P(1,\bm{r})
  =  \frac{n_{0} \rho_{0}  k_{B} T}{2}
  \int d\bm{r} \, \left[ \frac{3\bm{r} \bm{r}}{R_{0}^{2}}
  -  \bm{1} \right] P(1,\bm{r})
\end{equation}

For the sparse network model, we define the stress tensor as follows,
instead of eq \eqref{stress_tensor_definition_single_chain}.
\begin{equation}
 \label{stress_tensor_definition_single_bond}
 \hat{\bm{\sigma}}(n) \equiv
  \begin{cases}
   \displaystyle \bar{n}_{\text{eff}} \rho_{0}
  \left[ \frac{\partial \bar{u}(\bm{r})}{\partial \bm{r}} \bm{r}
   - k_{B} T \bm{1} \right] & (n = 1) \\
   0 & (n = 0)
  \end{cases}
\end{equation}
Here $\bar{n}_{\text{eff}}$ is the effective number of elastically
active superbridges per
micellar core, and $\bar{n}_{\text{eff}} \rho_{0}$ represents the
effective number density of superbridges. The relation between
$\bar{n}_{\text{eff}}$ and $n_{0}$ is generally not simple, but
physically $\bar{n}_{\text{eff}}$ should be a monotonically increasing
function of the polymer concentration and it should approach $n_{0} / 2$ in the high
concentration limit.

Although there are qualitative differences between the dense and sparse
network models, their equilibrium probability distributions
or dynamic equations are apparently similar. Thus, in the followings, we mainly
describe the detailed derivations and analyses for the dense network model, which is
simpler than the sparse network model. The results for sparse network model are
obtained in a similar way, and we just show the results without detailed derivations.

\subsubsection{Shear Relaxation Modulus}
\label{rheological_properties_shear_relaxation_modulus}


Here we calculate the shear relaxation modulus $G(t)$ of the dense
network model. We assume that the system is in equilibrium at $t < 0$, and a small step shear
deformation is applied at $t = 0$.
For convenience, we rewrite the Fokker-Planck operators (in eq
\eqref{fokker_planck_operator_equilibrium_single_chain}) as
\begin{equation}
 \label{fokker_planck_operator_equilibrium_single_chain_step_strain}
 \mathcal{L}(n,t) P(\bm{r}) = \mathcal{L}_{0}(n) P(\bm{r})
  - \frac{\partial}{\partial r_{x}} \cdot \left[ \gamma \delta(t) r_{y} P(\bm{r}) \right] 
\end{equation}
where $\mathcal{L}_{0}(n)$ is the Fokker-Planck operator in absence of
flow (in equilibrium).
In eq
\eqref{fokker_planck_operator_equilibrium_single_chain_step_strain},
we have assumed that the shear flow direction
and the shear gradient direction are $x$ and $y$, respectively, and
$\gamma$ is the shear strain which is
assumed to be sufficiently small. The distribution functions can be
expanded into the power 
series of $\gamma$ and the $O(\gamma^{2})$ terms can be dropped when
we calculate the shear relaxation modulus (because it is a linear
response function\cite{Evans-Morris-book}).

From eqs \eqref{dynamic_equation_phi_single_chain} and
\eqref{fokker_planck_operator_equilibrium_single_chain_step_strain}, the
distribution functions at $t > 0$ can be 
calculated as follows.
\begin{equation}
 \label{phi_response_step_strain}
 \begin{split}
   \Phi(\bm{r},t)
  & = \Phi_{\text{eq}}(\bm{r}) - e^{t \mathcal{L}_{0}(0)} 
  \frac{\partial}{\partial r_{x}}
  \left[ \gamma r_{y} \Phi_{\text{eq}}(\bm{r}) \right] \\
  & = \Phi_{\text{eq}}(\bm{r}) + \frac{\gamma}{k_{B} T} e^{t \mathcal{L}_{0}(0)} 
  \left[ \frac{\partial v(\bm{r})}{\partial r_{x}}
  r_{y} \Phi_{\text{eq}}(\bm{r}) \right]
 \end{split}
\end{equation}
The dynamic equation for $P(0,t)$ (eq \eqref{dynamic_equation_p0_single_chain}) can be rewritten as follows.
\begin{equation}
 \label{p0_response_step_strain_differential_form}
  \frac{\partial P(0,t)}{\partial t}
  = \frac{1}{\tau_{0}} [1 - P(0,t)]
  -  \frac{\xi}{\tau_{0}} \bigg[ \int d\bm{r} \, e^{- 3 \bm{r}^{2} / 2 R_{0}^{2}}
 \Phi(\bm{r},t) \bigg] P(0,t)
\end{equation}
The effect of the step shear on $P(0,t)$ appears in the second term in the right
hand side of eq \eqref{p0_response_step_strain_differential_form} only
through $\Phi(\bm{r},t)$. However, from the symmetry, 
the integral over $\Phi(\bm{r},t)$ becomes
\begin{equation}
 \begin{split}
  & \int d\bm{r} \, e^{- 3 \bm{r}^{2} / 2 R_{0}^{2}}
  \Phi(\bm{r},t) \\
  & =
 \int d\bm{r} \, e^{- 3 \bm{r}^{2} / 2 R_{0}^{2}}
 \Phi_{\text{eq}}(\bm{r}) 
  + \frac{\gamma}{k_{B} T} \int d\bm{r} \, e^{- 3 \bm{r}^{2} / 2 R_{0}^{2}}
  \left[ e^{t \mathcal{L}_{0}(0)} 
  \left[ \frac{\partial v(\bm{r})}{\partial r_{x}}
  r_{y} \Phi_{\text{eq}}(\bm{r}) \right] \right] \\
  & =
 \int d\bm{r} \, e^{- 3 \bm{r}^{2} / 2 R_{0}^{2}}
 \Phi_{\text{eq}}(\bm{r})
 \end{split}
\end{equation}
This means that a small deformation does not affect $P(0,t)$.
\begin{equation}
 \label{p0_response_step_strain}
 P(0,t) = P_{\text{eq}}(0)
\end{equation}


From eqs \eqref{dynamic_equation_p1_single_chain},
\eqref{phi_response_step_strain} and \eqref{p0_response_step_strain}, we
have the following expression for $P(1,\bm{r},t)$.
\begin{equation}
 \label{p1_response_step_strain}
  \begin{split}
  P(1,\bm{r},t)
  & = P_{\text{eq}}(1,\bm{r}) 
  + \frac{\gamma}{k_{B} T} e^{t [\mathcal{L}_{0}(1) - 1 /
  \tau_{0}]}
  \left[ \left[ \frac{3 k_{B} T r_{x}}{R_{0}^{2}}
   +  \frac{\partial v(\bm{r})}{\partial r_{x}} \right]
   r_{y} P_{\text{eq}}(1,\bm{r}) \right]
  \\
  & \qquad + \frac{\gamma \xi P_{\text{eq}}(0)}{k_{B} T \tau_{0}} \int_{0}^{t} dt' \, e^{(t - t') [\mathcal{L}_{0}(1) - 1 /
  \tau_{0}]}
  \left[ e^{- 3 \bm{r}^{2} / 2 R_{0}^{2}}
  e^{t' \mathcal{L}_{0}(0)} 
  \left[ \frac{\partial v(\bm{r})}{\partial r_{x}}
  r_{y} \Phi_{\text{eq}}(\bm{r}) \right] \right]
  \end{split}
\end{equation}
The shear relaxation modulus is calculated from the average stress
tensor as $G(t) \equiv \sigma_{xy}(t) / \gamma$.
Substituting eq \eqref{p1_response_step_strain} into
eq \eqref{stress_tensor_definition_single_chain}, we finally have the
expression of the shear relaxation modulus $G(t)$.
\begin{equation}
 \label{shear_relaxation_modulus_explicit}
 \begin{split}
  G(t)
  & = \frac{n_{0} \rho_{0}}{2} \int d\bm{r} \, \frac{3 r_{x} r_{y}}{R_{0}^{2}}
  \bigg[ e^{t [\mathcal{L}_{0}(1) - 1 /
  \tau_{0}]}
  \left[ \left[ \frac{3 k_{B} T r_{x}}{R_{0}^{2}}
  + \frac{\partial v(\bm{r})}{\partial r_{x}} \right] r_{y}
  P_{\text{eq}}(1,\bm{r}) \right] \\
  & \qquad + \frac{\xi P_{\text{eq}}(0)}{\tau_{0}} \int_{0}^{t} dt' \, e^{(t - t') [\mathcal{L}_{0}(1) - 1 /
  \tau_{0}]}
  \left[  e^{- 3 \bm{r}^{2} / 2 R_{0}^{2}}
  e^{t' \mathcal{L}_{0}(0)} 
  \left[ \frac{\partial v(\bm{r})}{\partial r_{x}}
  r_{y} \Phi_{\text{eq}}(\bm{r}) \right] \right]  \bigg]
 \end{split}
\end{equation}
Intuitively, this result can be interpreted as follows.
The first term represents the relaxation by the thermal motion and
reconstruction of the network (bridges). The second term represents the coupling between the
spatial correlation of micellar cores and the bridge construction. That
is, if the structure is deformed by the applied shear deformation, newly
constructed bridge chains are also deformed. Such a coupling gives
non-trivial relaxation modes. The second term in the right hand side of
eq \eqref{shear_relaxation_modulus_explicit} is
missing in most of previous models. Eq
\eqref{shear_relaxation_modulus_explicit} reduces to a single Maxwellian
form in the limit of $\xi \to 0$ and $\mathcal{L}_{0} \to 0$.
This limit is, however, physically unreasonable because the network is
no longer dense: The functionality becomes very small for $\xi \to 0$ and the
network becomes sparse. Also, the correlation between micellar cores is
not small and thus $\mathcal{L}_{0}$ is not negligible.

For a sparse network, the condition
considered above turn to be reasonable.
Generally the effective activity (fugacity)
$\bar{\xi}$ is small, and the Fokker-Planck operator in absence of
flow is negligible compared with other contributions ($\bar{\mathcal{L}}_{0}(1)
\to 0$). The characteristic network size becomes larger than the
characteristic interaction range between micellar cores, and thus the
effects of spatial correlation between micelles become small.
Therefore, here we consider the shear relaxation modulus
of the sparse network model for the cases where $\bar{\xi} \ll 1$,
$\bar{\mathcal{L}}_{0}(1) \approx 0$, and $v(\bm{r}) / k_{B} T \ll 1$. The
effective potential $\bar{u}(\bm{r})$  and the Fokker-Planck operator simply
become
\begin{align}
 & \label{effective_elastic_potential_single_bond}
 \bar{u}(\bm{r}) \approx \frac{3 k_{B} T}{2 m R_{0}^{2}} \bm{r}^{2} \\
 & \label{fokker_planck_operator_l1_single_bond}
 \bar{\mathcal{L}}(1,t) P(\bm{r}) \approx
  - \frac{\partial}{\partial r_{y}} [\gamma r_{x} \delta(t) P(\bm{r})]
\end{align}
and the shear relaxation modulus and characteristic modulus have
simple forms.
\begin{align}
 & \label{shear_relaxation_modulus_single_bond}
   G(t) \approx G_{0} e^{- m t / \tau_{0}} \\
 & \label{characteristic_modulus_single_bond}
   G_{0}
   \approx \bar{n}_{\text{eff}} \rho_{0} k_{B} T  \int d\bm{r} \,
   \frac{3 r_{x} r_{y}^{2}}{m R_{0}^{2}}
   \left[ \frac{3 r_{x}}{m R_{0}^{2}}
   + \frac{\partial v(\bm{r})}{\partial r_{x}} \right]
   \bar{P}_{\text{eq}}(1,\bm{r})
\end{align}
This result seems to be in harmony with the experimentally observed
single Maxwellian type relaxation behavior of the telechelic polymer 
solutions with intermediate concentrations, in which sparse networks
of superbridges are expected.

\subsubsection{Shear Viscosity and First Normal Stress Coefficient}


Under fast flows, telechelic polymer solutions show some
nonlinear rheological properties. Here we consider the shear viscosity
and first normal stress coefficient under shear flow.
For the dense network model, they are defined as
\begin{align}
 & \label{steady_shear_viscosity_definition}
 \eta(\dot{\gamma}) \equiv \frac{1}{\dot{\gamma}} \int d\bm{r} \,
 \hat{\sigma}_{xy}(1) P_{\text{ss}}(1,\bm{r}) \\
 & \label{steady_first_normal_stress_coefficient_definition}
 \Psi_{1}(\dot{\gamma}) \equiv \frac{1}{\dot{\gamma}^{2}} \int d\bm{r} \,
 [\hat{\sigma}_{xx}(1) - \hat{\sigma}_{yy}(1)] P_{\text{ss}}(1,\bm{r})
\end{align}
where $\dot{\gamma}$ is the shear rate, and the shear flow and shear
gradient directions are set to the $x$ and $y$ directions, as before.
The quantities with the subscript ``ss'' represent 
the steady state quantities under shear. Experimentally, both $\eta(\dot{\gamma})$
and $\Psi_{1}(\dot{\gamma})$ are known to exhibit nonlinear behavior.
(The nonlinear behavior of $\eta(\dot{\gamma})$
and $\Psi_{1}(\dot{\gamma})$ depend on various factors
such as the polymer concentration\cite{Suzuki-inpreparation}.)

The steady state micellar core distribution function is
given as the solution of the following equation.
\begin{equation}
  \label{steady_state_distribution_function_phi}
\mathcal{L}_{\text{ss}}(0) \Phi_{\text{ss}}(\bm{r}) = 0
\end{equation}
We cannot obtain the explicit form of $\Phi_{\text{ss}}(\bm{r})$ because we have not
specified the Fokker-Planck operator. (Even if the explicit form of $\mathcal{L}_{\text{ss}}(0)$ is given,
generally it is still difficult to obtain the explicit form of $\Phi_{\text{ss}}(\bm{r})$.)
However, some qualitative properties of $\Phi_{\text{ss}}(\bm{r})$ can
be argued without such detailed information. An important and general property
of $\Phi_{\text{ss}}(\bm{r})$ is that it is anisotropic. This is because the
operator $\mathcal{L}_{\text{ss}}(0)$ contains the advection
term, which is intrinsically anisotropic. If the mobility tensor becomes anisotropic, the anisotropy will be
enhanced.

The steady state probability distributions $P_{\text{ss}}(0)$ and
$P_{\text{ss}}(1,\bm{r})$ can be formally expressed in terms of $\Phi_{\text{ss}}(\bm{r})$.
\begin{equation}
  P_{\text{ss}}(0) = \left[ 1 + \xi \int d\bm{r} \,
   e^{- 3 \bm{r}^{2} / 2 R_{0}^{2}}
 \Phi_{\text{ss}}(\bm{r}) \right]^{-1}
\end{equation}
\begin{equation}
 \label{steady_state_distribution_p1_asymptotic}
 P_{\text{ss}}(1,\bm{r}) = P_{\text{ss}}(0) \,
  \xi \sum_{k = 0}^{\infty} \tau_{0}^{k} \mathcal{L}_{\text{ss}}^{k}(1)
  [  e^{- 3 \bm{r}^{2} / 2 R_{0}^{2}}
 \Phi_{\text{ss}}(\bm{r}) ]
\end{equation}
Thus $P_{\text{ss}}(1,\bm{r})$ is also anisotropic under
shear. Moreover, the operator $\mathcal{L}_{\text{ss}}(1)$
modulates the anisotropy in a rather complicated way. Therefore,
the effect of the shear flow on $P_{\text{ss}}(1,\bm{r})$ is not
trivial even in this simple case.

From eqs \eqref{steady_shear_viscosity_definition},
\eqref{steady_first_normal_stress_coefficient_definition}
and \eqref{steady_state_distribution_p1_asymptotic},
the shear viscosity and the first normal stress coefficient become
\begin{equation}
 \label{steady_shear_viscosity_explicit}
  \eta(\dot{\gamma})
  = \frac{n_{0} \rho_{0} \xi k_{B} T}{2 \dot{\gamma}}
  P_{\text{ss}}(0)
  \int d\bm{r} \,
  \frac{3 r_{x} r_{y}}{R_{0}^{2}}
 \sum_{k = 0}^{\infty} \tau_{0}^{k} \mathcal{L}_{\text{ss}}^{k}(1)
  [  e^{- 3 \bm{r}^{2} / 2 R_{0}^{2}}
 \Phi_{\text{ss}}(\bm{r}) ]
\end{equation}
\begin{equation}
 \label{steady_first_normal_stress_coefficient_explicit}
  \Psi_{1}(\dot{\gamma})
  = \frac{n_{0} \rho_{0} \xi k_{B} T}{2 \dot{\gamma}^{2}} 
   P_{\text{ss}}(0) \int d\bm{r} \, \frac{3 (r_{x}^{2} -  r_{y}^{2})}{R_{0}^{2}}
 \sum_{k = 0}^{\infty} \tau_{0}^{k} \mathcal{L}_{\text{ss}}^{k}(1)
  [  e^{- 3 \bm{r}^{2} / 2 R_{0}^{2}}
 \Phi_{\text{ss}}(\bm{r}) ]
\end{equation}
In general, both $\Phi_{\text{ss}}(\bm{r})$ and
$\mathcal{L}_{\text{ss}}(1)$ are anisotropic, and thus the dependence
of $\eta(\dot{\gamma})$ and $\Psi_{1}(\dot{\gamma})$ on $\dot{\gamma}$
is not simple.

For the sparse network model, we can approximate the Fokker-Planck
operator as $\bar{\mathcal{L}}(1,t)
\approx - \dot{\gamma} r_{y} (\partial / \partial r_{x})$.
Then we have the following expressions for the shear
viscosity and first normal stress coefficient.
\begin{equation}
 \label{steady_state_distributuion_single_effective_bond}
 \bar{P}_{\text{ss}}(0)
  \approx 
  \bigg[ 1 + \bar{\xi}
  \int d\bm{r} \, e^{- 3 \bm{r}^{2} / 2 m R_{0}^{2}}
  \widetilde{\Phi^{[m]}_{\text{ss}}}(\bm{r}) \bigg]^{-1} 
\end{equation}
\begin{equation}
 \label{steady_shear_viscosity_single_bond}
 \eta(\dot{\gamma})
  \approx \frac{\bar{n}_{\text{eff}} \rho_{0} \bar{\xi} k_{B} T}{
  \dot{\gamma}} \bar{P}_{\text{ss}}(0)
 \int d\bm{r} \, 
  \frac{3}{m R_{0}^{2}} \left[  r_{x} r_{y}
  + \left(\frac{\tau_{0} \dot{\gamma}}{m}\right) r_{y}^{2} \right]
  e^{- 3 \bm{r}^{2} / 2 m R_{0}^{2}}
  \widetilde{\Phi_{\text{ss}}^{[m]}}(\bm{r}) 
\end{equation}
\begin{equation}
 \label{steady_first_normal_stress_coefficient_single_bond}
  \begin{split}
   \Psi_{1}(\dot{\gamma})
   & \approx \frac{\bar{n}_{\text{eff}} \rho_{0} \bar{\xi} k_{B} T}{
  \dot{\gamma}^{2}}
 \bar{P}_{\text{ss}}(0) \int d\bm{r} \, \frac{3}{m R_{0}^{2}}
  \bigg[ r_{x}^{2} - r_{y}^{2} \\
   & \qquad + 2 \left(\frac{\tau_{0} \dot{\gamma}}{m}\right) r_{x} r_{y} 
  + 2 \left(\frac{\tau_{0} \dot{\gamma}}{m}\right)^{2} r_{y}^{2}
   \bigg] e^{- 3 \bm{r}^{2} / 2 m R_{0}^{2}}
  \widetilde{\Phi_{\text{ss}}^{[m]}}(\bm{r})
  \end{split}
\end{equation}
Eqs \eqref{steady_shear_viscosity_single_bond} and
\eqref{steady_first_normal_stress_coefficient_single_bond} may seem to
be somehow similar to eqs \eqref{steady_shear_viscosity_explicit} and
\eqref{steady_first_normal_stress_coefficient_explicit}, and one may consider
that the nonlinearities in the sparse network model are rather
weak. Indeed, the origin of the nonlinearities itself is qualitatively the same
as in the case of the dense network model. (The nonlinearities mainly come from
the nonlinear and anisotropic dependence of $\Phi_{\text{ss}}(\bm{r})$
on shear flow.)
However, we should recall that
$\widetilde{\Phi_{\text{ss}}^{[m]}}(\bm{r})$ is expressed in terms of convolutions of
$\Phi_{\text{ss}}(\bm{r})$  (eq \eqref{tilde_convolutions_definition}),
and thus its nonlinearity will be enhanced.
Therefore, the shear rate dependence deduced from
eqs \eqref{steady_shear_viscosity_single_bond} and
\eqref{steady_first_normal_stress_coefficient_single_bond} 
can be stronger than that in the dense network model.

Here, we examine the shear thickening and thinning behavior.
We expand the function
$\widetilde{\Phi_{\text{ss}}^{[m]}}(\bm{r})$ into a Hermite polynomial
series, following the previous work\cite{Suzuki-Uneyama-Inoue-Watanabe-2012}.
\begin{equation}
 \label{phi_tilde_hermite_expansion}
\begin{split}
 e^{- 3 \bm{r}^{2} / 2 m R_{0}^{2}} \widetilde{\Phi_{\text{ss}}^{[m]}}(\bm{r})
 & = 
 \left(\frac{3}{2 \pi m R_{0}^{2}}\right)^{3/2} e^{- 3 \bm{r}^{2} / 2 m R_{0}^{2}}
   \sum_{i,j,k = 0}^{\infty} B_{i,j,k}(\dot{\gamma})
  2^{- (i + j + k) / 2} \\
 & \qquad \times H_{i}\left( \frac{\sqrt{3} r_{x}}{\sqrt{2 m} R_{0}} \right)
  H_{j}\left( \frac{\sqrt{3} r_{y}}{\sqrt{2 m} R_{0}} \right)
  H_{k}\left( \frac{\sqrt{3} r_{z}}{\sqrt{2 m} R_{0}} \right)
\end{split}
\end{equation}
Here $H_{i}(x)$ is the $i$-th order Hermite polynomial\cite{Abramowitz-Stegun-book}, and
$\lbrace B_{i,j,k}(\dot{\gamma}) \rbrace$ are the expansion
coefficients.
Substituting eq \eqref{phi_tilde_hermite_expansion} into eqs
\eqref{steady_state_distributuion_single_effective_bond}-\eqref{steady_first_normal_stress_coefficient_single_bond},
we find that the shear viscosity and first normal stress coefficient are
expressed in terms of $\lbrace B_{i,j,k}(\dot{\gamma}) \rbrace$ as
\begin{equation}
 \label{steady_shear_viscosity_single_bond_expanded}
  \eta(\dot{\gamma})
  \approx \frac{\bar{n}_{\text{eff}} \rho_{0} \bar{\xi} k_{B} T}{
  \dot{\gamma} [ 1 + \bar{\xi} B_{0,0,0}(\dot{\gamma}) ]}
  \left[  B_{1,1,0}(\dot{\gamma})
  + \left(\frac{\tau_{0} \dot{\gamma}}{m}\right)
  [ B_{0,0,0}(\dot{\gamma}) + 2 B_{0,2,0}(\dot{\gamma}) ] \right]
\end{equation}
\begin{equation}
 \label{steady_first_normal_stress_coefficient_single_bond_expanded}
  \begin{split}
    \Psi_{1}(\dot{\gamma})
   & \approx \frac{2 \bar{n}_{\text{eff}} \rho_{0} \bar{\xi} k_{B} T}{
  \dot{\gamma}^{2} [ 1 + \bar{\xi} B_{0,0,0}(\dot{\gamma}) ]}
  \bigg[ B_{2,0,0}(\dot{\gamma}) -
   B_{0,2,0}(\dot{\gamma}) \\
   & \qquad + \left(\frac{\tau_{0} \dot{\gamma}}{m}\right) B_{1,1,0}(\dot{\gamma})
  + \left(\frac{\tau_{0} \dot{\gamma}}{m}\right)^{2} 
   [B_{0,0,0}(\dot{\gamma}) + 2 B_{0,2,0}(\dot{\gamma})]
   \bigg]
  \end{split}
\end{equation}
From eqs \eqref{steady_shear_viscosity_single_bond_expanded} and
\eqref{steady_first_normal_stress_coefficient_single_bond_expanded}, it
is clear that the shear thickening and/or thinning are determined by the
shear rate dependence of four coefficients,
$B_{0,0,0}(\dot{\gamma})$, $B_{1,1,0}(\dot{\gamma})$, $B_{2,0,0}(\dot{\gamma})$,
and $B_{0,2,0}(\dot{\gamma})$.
$\eta(\dot{\gamma})$ and $\Psi_{1}(\dot{\gamma})$ depend on these
coefficients differently.
Also, the $\dot{\gamma}$-dependence of these
coefficients is generally not simple. As a result of the competition
between these coefficients,
several different types of nonlinear behavior can be reproduced.
Because the anisotropy is enhanced in the case of the sparse network
model, we expect that the sparse network model can exhibit variety of
shear thickening and/or thinning behavior.
(We can further analyze nonlinear behavior by expanding
$\lbrace B_{i,j,k}(\dot{\gamma}) \rbrace$ into power series of $\dot{\gamma}$\cite{Suzuki-Uneyama-Inoue-Watanabe-2012},
although we do not show details here.)

\section{Discussion}
\label{discussion}

\subsection{Shear Relaxation Modulus}

The shear relaxation modulus of the dense network model (eq
\eqref{shear_relaxation_modulus_explicit}) does not reduce to
the single Maxwellian relaxation as shown in Sec.~\ref{rheological_properties_shear_relaxation_modulus}.
As we mentioned, the conditions for the single Maxwellian relaxation
($g_{\text{eq}}(\bm{r}) \approx 1$ and $\xi \ll 1$ (or $\bar{\xi} \ll
1$)) are satisfied if the concentration of micellar cores 
is low and the network is sparse.
Experimental data shown in this work have suggested that the crossover
concentration of our HEUR 
solutions is about $c_{c} \approx 4\text{wt\%}$ (cf. Figures
\ref{heur_concentration_dependence_rescaled} and \ref{heur_tau0_g0_concentration_dependence}).
For $c_{c} \lesssim 4\text{wt\%}$ we can employ the sparse network model
and thus the single Maxwellian relaxation is reproduced. For $c_{c}
\gtrsim 4\text{wt\%}$, we should employ the dense network model.

The expression of the shear relaxation modulus in the dense network
model is not so simple.
To make the
expression analytically tractable, we replace the Fokker-Planck operators in eq 
\eqref{shear_relaxation_modulus_explicit} by constants, as
$\mathcal{L}_{0}(n) \to - 1 / \tilde{\tau}(n)$. Such a
replacement corresponds to an approximation for the operators by their
characteristic eigenvalues. With this approximate replacement, eq
\eqref{shear_relaxation_modulus_explicit} reduces to a simple form.
\begin{equation}
 \label{shear_relaxation_modulus_explicit_constant_relaxation_time_approximation}
 \begin{split}
  G(t)
  & \approx
  G_{0} e^{- t [ 1 / \tilde{\tau}(1) + 1 / \tau_{0}]}
  + G_{1} \frac{e^{- t  / \tilde{\tau}(0)}
  - e^{- t [ 1 / \tilde{\tau}(1) + 1 / \tau_{0}]}}{ 1 + \tau_{0} / \tilde{\tau}(1) - \tau_{0}  / \tilde{\tau}(0)}
 \end{split}
\end{equation}
$G_{1}$ is a sort of characteristic modulus defined as follows.
\begin{equation}
 G_{1} \equiv \frac{n_{0} \rho_{0}  \xi
  P_{\text{eq}}(0) }{2} \int d\bm{r} \,
 \frac{3 r_{x} r_{y}^{2}}{R_{0}^{2}}
 \frac{\partial v(\bm{r})}{\partial r_{x}}
  e^{- 3 \bm{r}^{2} / 2 R_{0}^{2}}
 \Phi_{\text{eq}}(\bm{r})
\end{equation}
($G_{1}$ is expected to be small compared with $G_{0}$.)
The first term in the right hand side of eq
\eqref{shear_relaxation_modulus_explicit_constant_relaxation_time_approximation}
is the single Maxwell type relaxation. The characteristic
relaxation time is given as $\tau_{0} \tilde{\tau}(1) / [\tau_{0} +
\tilde{\tau}(1)]$. (This relaxation time is longer than the intrinsic time
$\tau_{0}$ and changes with the polymer
concentration.)
The second term in the right hand side of eq
\eqref{shear_relaxation_modulus_explicit_constant_relaxation_time_approximation}
represents to the deviation from a single Maxwellian form, and it
originates from the coupling between the spatial distribution of
micellar cores and the bridge construction.
The characteristic relaxation times $\tilde{\tau}(0)$ and
$\tilde{\tau}(1)$ are expected to be larger than $\tau_{0}$. In the
short and long time limit, the deviation term asymptotically becomes
\begin{equation}
 \label{shear_relaxation_modulus_deviation_term_asymptotic}
 \frac{e^{- t  / \tilde{\tau}(0)}
  - e^{- t [ 1 / \tilde{\tau}(1) + 1 / \tau_{0}]}}{ 1 + \tau_{0} /
  \tilde{\tau}(1) - \tau_{0}  / \tilde{\tau}(0)} \to
 \begin{cases}
 t / \tau_{0} + O\left( (t / \tau_{0})^{2} \right) & (t \ll \tau_{0}) \\
  \displaystyle  \frac{e^{- t  / \tilde{\tau}(0)}}{ 1 + \tau_{0} /
  \tilde{\tau}(1) - \tau_{0}  / \tilde{\tau}(0)} & (t \gg \tau_{0})
 \end{cases}
\end{equation}
Intuitively, the deviation term slightly decelerates the relaxation in the short
time scale and add a slow and weak relaxation mode in the long time scale. As a result, the relaxation spectrum
becomes broader than the single Maxwellian spectrum.

\begin{figure}[h]
 \centering
 {\includegraphics[width=0.9\linewidth,clip]{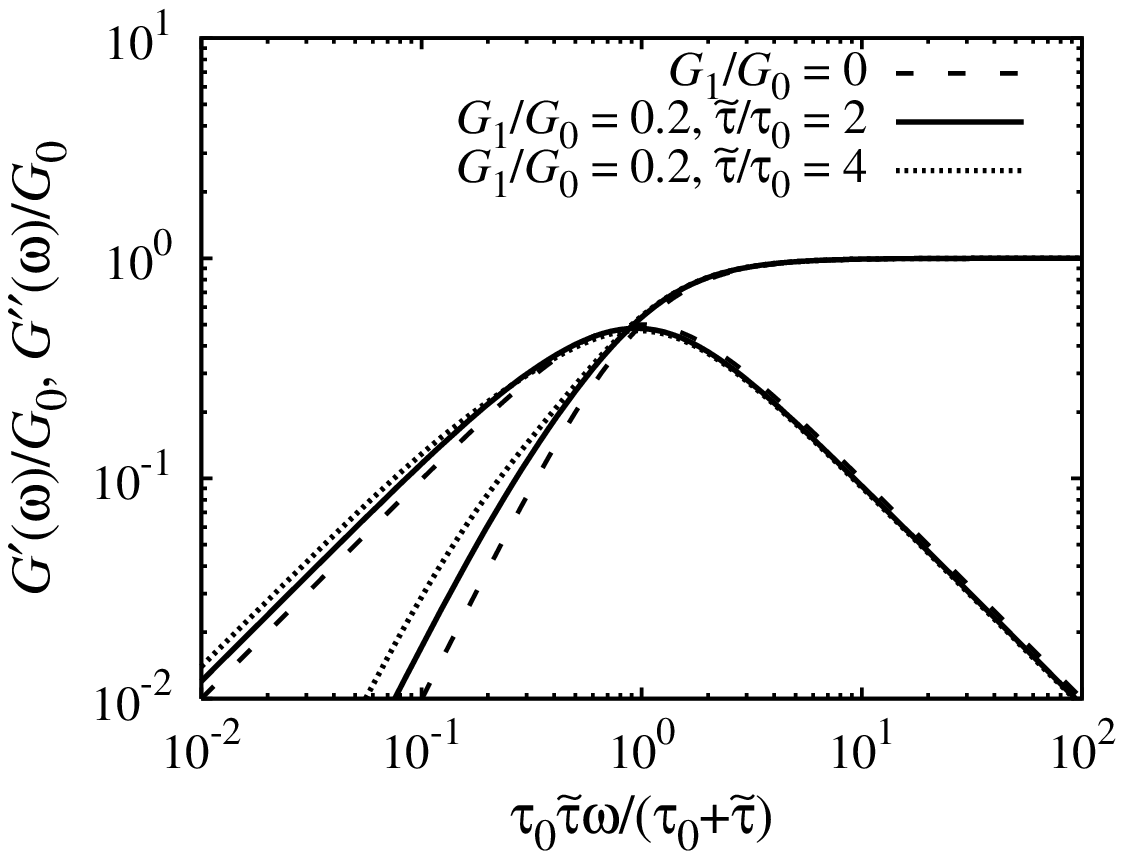}}
 \caption{\label{storage_and_loss_moduli_dense_network}}
\end{figure}

As an example, we show the storage and loss moduli, $G'(\omega)$ and
$G''(\omega)$ calculated from eq
\eqref{shear_relaxation_modulus_explicit_constant_relaxation_time_approximation}
for some parameter values
in Figure
\ref{storage_and_loss_moduli_dense_network}.
The deviation from the single Maxwellian behavior is clearly observed.
Similar trend can be observed in the
experimental data, Figure \ref{heur_concentration_dependence_rescaled}.
We expect that this is because our dense network model captures the essential
physics qualitatively.
However, we should notice that the estimate shown here is rough and cannot be
quantitatively compared with experimental data. In reality, the Fokker-Planck
operators give multiple relaxation times and the shear relaxation
modulus (or the storage and loss moduli) will be much broader.
Here it may be worth mentioning that the deviation from the single Maxwellian is already
predicted in the transient network model by Tanaka and
Koga\cite{Tanaka-Koga-2006}. However, in their 
model, the storage and loss moduli data become
sharper (in the mode distribution), whereas our model and the experimental data show the
broadening. This difference could result from the lack of the correlation
between micellar cores in the Tanaka-Koga model.
%


\subsection{Concentration Dependence of Characteristic Time and Modulus}

Experimental data show clear concentration dependence of the
characteristic relaxation time $\tau$ and modulus
$G_{0}$. In the single chain transient network models without spatial
correlation effect and the network functionality, the experimentally observed concentration dependence
cannot be reproduced.

We consider the dependence of the characteristic
relaxation time $\tau$ in the conventional single chain models.
The characteristic relaxation time $\tau$ is just the same as the
average bridge destruction time $\tau_{0}$, $\tau = \tau_{0}$.
$\tau_{0}$ is considered to be determined by the characteristic relaxation time of a
polymer chain (such as the Rouse and Zimm times) and the
activation energy (energy barrier).
Naively, both of them are almost independent
of the polymer concentration $c$, and we can roughly estimate as
$\tau \propto c^{0}$.
This estimate for the concentration dependence of $\tau$
is not consistent with experimental data. (Although it may be
possible to remedy the model by introducing some concentration
dependence to parameters such as $\tau_{0}$, such an
ad-hoc approach is not always physically reasonable.)

In our models, the concentration dependence of $\tau$ is not that
simple. For the dense network model, due to the coupling between the
association-dissociation dynamics and the dynamics of micellar cores, the
characteristic relaxation time has been estimated to be $\tau \approx \tau_{0} \tilde{\tau}(1)
/ [\tau_{0} + \tilde{\tau}(1)]$ (see eq
\eqref{shear_relaxation_modulus_explicit_constant_relaxation_time_approximation}). This
relaxation time can depend on $c$ through $\tilde{\tau}(1)$.
For the sparse network model, from eq
\eqref{shear_relaxation_modulus_single_bond} the characteristic relaxation time is
given as $\tau = \tau_{0} / m$, which can also depend on $c$ because
$m$ is a function of $c$. Besides, the $c$-dependence of $\tau$
of the dense and sparse network models is expected to be different
(because of the difference between the $c$-dependence of $\tilde{\tau}(1)$ and $m$).
This is qualitatively consistent with the experimental results
(Figure \ref{heur_tau0_g0_concentration_dependence}(a)).


\subsection{Fast Mode}

The storage and loss moduli data of the HEUR solutions exhibit the fast mode
at the high frequency region. Our models do not take into
account the local relaxation dynamics of polymer chains and thus they
cannot reproduce the fast mode. Nonetheless, we can estimate how the
fast mode behaves for the dense and spare network cases.

In micellar network structures, chain ends of associative polymers are
connected each other and thus the relaxation mode distributions are given
as the eigenmode distributions of networks. This situation is similar to
the Rouse model, where the linear viscoelasticity is described by
eigenmodes of linearly connected springs.
For the sparse network, we
expect that the fast mode mainly reflects the relaxation of a
superbridge structure. In analogy to the Rouse model,
the relaxation time depends on the number of chains per superbridge $m$
as $\tau_{\text{fast}} \propto m^{2} \tau_{\text{PEO}}$. Here
$\tau_{\text{PEO}}$ is the relaxation time of a free PEO chain. Thus the relaxation time of the
fast mode would be longer than the relaxation time of the PEO
solution. Moreover, the viscosity depends on $m$ as $\eta_{\text{fast}}
\propto \nu_{\text{eff}} m^{2} \propto m$, with $\nu_{\text{eff}}
\propto m^{-1}$ being
the effective number density of superbridges. Thus $G''(\omega)$ of the
HEUR solution is expected to be larger than the PEO solution. Although
the estimate above is very rough, the trend is consistent with
experimental data (Figure \ref{heur_1wt_master_curves_slow}).

For the dense network, the situation is qualitatively different.
In the dense network, polymer chains are connected in a rather
complicated way. The eigenmode distributions will be much broader than
the case of the sparse network, and the relaxation time will be much longer.
As a result, the fast mode of the dense network will be broader
than the sparse network. Also, the relaxation time of the fast mode will
be longer than one of the sparse network. If we assume that
networks are fractal structures, the storage and loss moduli are
described by the power law (just as in cases of gels\cite{Winter-Chambon-1986}).
The power law exponent is expected to decrease as the concentration
increases and the network becomes denser (well-percolated). This is consistent with experimental data
shown in Figure \ref{heur_concentration_dependence}. Therefore we can
again conclude that, for both the spare and dense networks, the fast mode
reflects the relaxation behavior of main chains in the connected
(associated) network structures.

\subsection{Nonlinear Elasticity and Stretch Dependent Reconstruction Rate}

Here we discuss the effects of the nonlinear elasticity and the
stretch dependent reconstruction rates in detail.
In particular, we focus on mechanisms for the shear thickening and
thinning, for which these factors have been assumed to be important.

In some of conventionally examined transient network
models, the shear thickening and
 thinning are explained as a result of the competition between two factors;
the nonlinear elasticity and the stretch dependent dissociation
rate\cite{Wang-1992a,Marrucci-Bhargava-Cooper-1933,Vaccaro-Marrucci-2000,Indei-Koga-Tanaka-2005,Indei-2007,Indei-2007a,Koga-Tanaka-2010}.
Under fast flow, the nonlinear elasticity increases the shear stress
whereas the stretch dependent dissociation decreases the number of
highly stretched chains and thus decreases the shear stress.
As we have shown, however, even without the nonlinear elasticity and the stretch
dependent reconstruction rates, some nonlinear rheological behavior can
be naturally reproduced. In our models, the shear stress is strongly
affected by the steady state distribution function
$\Phi_{\text{ss}}(\bm{r})$ given by eq \eqref{steady_state_distribution_function_phi}.
Thus, the structural anisotropy can be the main mechanism
giving rise to the shear thickening and/or thinning 
in actual telechelic polymer solutions.

This implies that the nonlinear elasticity and the stretch dependent
dissociation rate are not always essential ingredients of transient
network type models. There are many different possible dynamic models which can
reproduce the same shear thickening and/or thinning behavior. Even
if a specific model reproduces the shear thickening and/or 
thinning behavior of a specific sample perfectly (with some parameter fittings),
it does not guarantee that the model is physically correct.
Actually, a recent experimental work showed that the effect of nonlinear elasticity
is negligibly small for typical HEUR solutions, based on a simple
energy balance argument\cite{Suzuki-Uneyama-Inoue-Watanabe-2012}.
Careful and systematic comparison of theoretical predictions and
experimental data are absolutely necessary for validating a theoretical model.

Some conventional models predict that the average bridge chain is
smaller, in spatial size, compared with
the average loop chain size and/or dangling chain size in equilibrium.
(The fact that the stretch dependent dissociation rate affects the
equilibrium chain statistics has already pointed by Tanaka and
Edwards\cite{Tanaka-Edwards-1992}.)
However, judging from molecular simulation
data\cite{Khalatur-Khokhlov-1996} or scattering data
\cite{Serero-Aznar-Porte-Berret-Calvet-Collet-Viguer-1998,Serero-Jacobsen-Berret-2000},
the average bridge chain size is larger than the loop/dangling chain size. Moreover,
if the average bridge chain size is much smaller than the loop or
dangling chain size, uniformly spanned networks cannot be formed.
Under such conditions, some simple stretch dependent dissociation models
will become physically odd.
This consideration suggests that the spatial
correlation (spatial structure) is an important factor determining the
rheological behavior of telechelic polymer solutions.
Of course, the stretch dependence may be important for some cases.
The reassociation process in telechelic
polymer solutions needs to be modeled with very careful considerations.

It is fair to mention that our models still lack various features
which may be important for some cases.
For example, the direct
bridge-to-bridge transition is not included in dynamic equations
\eqref{dynamic_equation_p1_single_chain} or \eqref{dynamic_equation_p1_single_bond}.
The bridge-to-bridge transition can contribute to several rheological
properties, especially when a network is dense and the number density of
micellar cores is high.
The higher order spatial correlation functions,
being important for such dense systems,
are not explicitly considered in our models. If we rigorously construct the dynamic equation for the two
body correlation in dense systems, generally we have the three body correlation function.
(This is qualitatively the same as so-called the BBGKY hierarchy in the
liquid state theory\cite{Hansen-McDonald-book}.) Also, the interactions
between bridge chains or micellar cores are considered in our models
only via the mean
field type approximation, which may not be always acceptable.
The polydispersity of molecular weights
or aggregation number are also ignored in our models.
These features are beyond the scope of
this work and are not further discussed here.
Elaborated
theoretical investigations as well as detailed molecular simulations
will be required to understand the molecular mechanisms in telechelic
polymer solutions.

\section{Conclusions}

We conducted linear viscoelastic measurements for HEUR solutions with
different polymer concentrations. We have shown that the HEUR solutions
exhibit fast and slow relaxations, and both of them depend on the
concentration. Our HEUR solutions exhibited different rheological behavior at
low and high concentrations.
These results suggest existence of two different concentration regimes for
telechelic polymer solutions.

To explain these experimental results,
we have considered single chain type transient network models for dense
and sparse networks, which take into account the information of spatial
correlations between micellar cores and the network functionality.
The spatial correlation largely affects the bridge construction dynamics in our models.
We considered the linear and nonlinear rheological properties of
the models, in absence of the conventionally considered mechanisms,
such as nonlinear elasticity and the stretch dependent
bridge destruction rate. The sparse network model reproduces the
well-known single Maxwellian type shear relaxation modulus for
solutions of low polymer concentrations. On the other hand, 
the dense network model gives non-single
Maxwellian type shear relaxation modulus, which is consistent with
experimental results for high concentration HEUR solutions.

For the steady shear viscosity and first normal stress
coefficients, both the dense and sparse network models
exhibit nonlinear behavior such as the shear thickening and thinning.
The nonlinear rheological behavior originates mainly from by
the steady state distribution function of micellar cores,
$\Phi_{\text{ss}}(\bm{r})$.
This is consistent with a recently proposed anisotropic bridge formation
model\cite{Suzuki-Uneyama-Inoue-Watanabe-2012}. Our models suggest that
the nonlinear rheological properties are also strongly affected by the
polymer concentration. The spatial correlation is one possible origin
of the nonlinear rheological behavior. The measurements and analyses of
nonlinear rheological properties for the HEUR solutions are now in progress, and
will be reported elsewhere\cite{Suzuki-inpreparation}.

\section*{Acknowledgment}

This work was supported by Grant-in-Aid (KAKENHI) for Young
Scientists B 22740273.

\appendix


\section{Anisotropic Bridge Formation Model}
\label{anisotropic_bridge_formation_model}

In this appendix, we briefly show the anisotropic bridge formation model
proposed in Ref \cite{Suzuki-Uneyama-Inoue-Watanabe-2012}.
We also show that the sparse network model formulated in this work 
reduces to the anisotropic bridge formation model under some conditions.

First we briefly show the anisotropic bridge formation model\cite{Suzuki-Uneyama-Inoue-Watanabe-2012}.
In this model, 
the number fraction of bridges (superbridges) is assumed to be constant
even under shear. We introduce the normalized probability distribution
for an end-to-end vector, $\psi(\bm{r},t)$. The dynamic equation under a steady
shear flow is described by a birth-death type master equation\cite{vanKampen-book}.
\begin{equation}
 \label{dynamic_equation_psi_anisotropic_bridge_formation_model}
 \frac{\partial \psi(\bm{r},t)}{\partial t}
  = - \dot{\gamma} r_{y} \frac{\partial \psi(\bm{r},t)}{\partial r_{x}} 
  + \frac{1}{\bar{\tau}_{0}} \phi(\bm{r},t)
  - \frac{1}{\bar{\tau}_{0}} \psi(\bm{r},t) 
\end{equation}
where we have set the shear flow and shear gradient directions to the $x$ and
$y$ directions, respectively. $\dot{\gamma}$ is the shear rate, $\bar{\tau}_{0}$ is
the characteristic bridge reconstruction time, and $\phi(\bm{r},t)$ is
the distribution function for a newly constructed bridge (superbridge).
$\phi(\bm{r},t)$ can be interpreted as a source function.
In general, $\phi(\bm{r},t)$ depends on the shear rate $\dot{\gamma}$, and 
the nonlinear rheological properties are mainly determined by $\phi(\bm{r},t)$.
The detailed dynamics of $\phi(\bm{r},t)$ does not need to be fully
specified. (In Ref \cite{Suzuki-Uneyama-Inoue-Watanabe-2012},
only some expansion coefficients at the steady state are utilized for the analysis.)
If the elastic potential of the end-to-end vector is given as a
harmonic form, the stress tensor can be expressed simply as
\begin{equation}
 \label{stress_tensor_anisotropic_bridge_formation_model}
 \bm{\sigma} = \nu k_{B} T \int d\bm{r}
  \left[ \frac{3 \bm{r} \bm{r}}{r_{0}^{2}}
  - \bm{1} \right]
  \psi(\bm{r})
\end{equation}
where $\nu$ is the number density of elastically active bridges and
$r_{0}$ is their average size. From
eqs \eqref{dynamic_equation_psi_anisotropic_bridge_formation_model} and
\eqref{stress_tensor_anisotropic_bridge_formation_model}, one can
analyze some rheological properties.

Now, we show that our sparse network model can reduce to the anisotropic bridge
formation model. Because the polymer concentration of the HEUR solution
used in Ref.~\cite{Suzuki-Uneyama-Inoue-Watanabe-2012} is
$1\text{wt\%}$, the experimental system is in the
sparse network regime as judged from the crossover
concentration in
Figures \ref{heur_concentration_dependence_rescaled} and
\ref{heur_tau0_g0_concentration_dependence} ($c_{c} \approx
4\text{wt\%}$).
We define the normalized bond vector distribution function as
\begin{equation}
 \label{anisotropic_bridge_formation_model_bridge_distribution}
 \psi(\bm{r},t) \equiv \frac{\bar{P}(1,\bm{r},t)}{1 -
 \bar{P}_{\text{eq}}(0)} 
\end{equation}
and the normalized source function as
\begin{equation}
 \label{anisotropic_bridge_formation_model_source_function}
 \phi(\bm{r},t) \equiv \frac{\bar{W}(1, \bm{r} | 0)}{\bar{W}(0 | 1, \bm{r})}
 \widetilde{\Phi^{[m]}}(\bm{r},t) \frac{\bar{P}_{\text{eq}}(0) }{1 -
 \bar{P}_{\text{eq}}(0)}
 = m \bar{\xi} e^{- \bar{u}(\bm{r}) / k_{B} T}
 \widetilde{\Phi^{[m]}}(\bm{r},t) \frac{\bar{P}_{\text{eq}}(0) }{1 -
 \bar{P}_{\text{eq}}(0)}
\end{equation}
Eq \eqref{anisotropic_bridge_formation_model_source_function} means that the source
function is determined from the distribution function of
micellar cores, $\Phi(\bm{r},t)$.
(At the steady state under flow, the source
function $\phi_{\text{ss}}(\bm{r})$ is determined by
$\Phi_{\text{ss}}(\bm{r})$ in eq
\eqref{steady_state_distribution_function_phi}. Namely, the steady state Fokker-Planck operator
$\mathcal{L}_{\text{ss}}(0)$ determines $\phi_{\text{ss}}(\bm{r})$.)
Substituting eqs
\eqref{anisotropic_bridge_formation_model_bridge_distribution} and
\eqref{anisotropic_bridge_formation_model_source_function} into eq
\eqref{dynamic_equation_p1_single_bond}, we have the dynamic equation
for $\psi(\bm{r},t)$.
\begin{equation}
 \label{dynamic_equation_psi_anisotropic_bridge_formation_model_by_sparse_network_model}
 \frac{\partial \psi(\bm{r},t)}{\partial t}
  = \bar{\mathcal{L}}(1,t) \psi(\bm{r},t)
  + \frac{1}{\bar{\tau}_{0}} \phi(\bm{r},t)
  - \frac{1}{\bar{\tau}_{0}} \psi(\bm{r},t) 
\end{equation}
where we have set $\bar{\tau}_{0} = \tau_{0} / m$.
Further, if we neglect the Brownian motion and the interaction between
micelles and set
$\bar{\mathcal{L}}(1,t) \psi(\bm{r}) = - \dot{\gamma} r_{y} \partial \psi(\bm{r}) /
\partial r_{x}$, eq
\eqref{dynamic_equation_psi_anisotropic_bridge_formation_model_by_sparse_network_model}
reduces to eq \eqref{dynamic_equation_psi_anisotropic_bridge_formation_model}.
Also, the stress tensor is given as
\begin{equation}
 \label{stress_tensor_anisotropic_bridge_formation_model_by_sparse_network_model}
 \bm{\sigma} = \bar{n}_{\text{eff}} \rho_{0} [1 -
 \bar{P}_{\text{eq}}(0)]
  \int d\bm{r} \, \left[ \frac{\partial \bar{u}(\bm{r})}{\partial \bm{r}}{\bm{r}}
  - k_{B} T \bm{1} \right] \psi(\bm{r})
\end{equation}
With the harmonic form of the effective potential (eq
\eqref{effective_elastic_potential_single_bond} with $r_{0} = \sqrt{m} R_{0}$), eq
\eqref{stress_tensor_anisotropic_bridge_formation_model_by_sparse_network_model}
becomes essentially the same as eq \eqref{stress_tensor_anisotropic_bridge_formation_model}.
Thus the sparse network
model formulated in this work reduces to the anisotropic bridge formation model.

However, it should be noticed that
the dynamics of the source function is not explicitly specified
in the anisotropic bridge formation
model\cite{Suzuki-Uneyama-Inoue-Watanabe-2012}. This means that we can
use eq 
\eqref{dynamic_equation_psi_anisotropic_bridge_formation_model} even
when the source function is not given by eq
\eqref{anisotropic_bridge_formation_model_source_function}.
There are many other possible molecular models which reduce to the
anisotropic bridge formation model.

\bibliographystyle{apsrev4-1}
\bibliography{telechelic}

\begin{thebibliography}{39}%
\makeatletter
\providecommand \@ifxundefined [1]{%
 \@ifx{#1\undefined}
}%
\providecommand \@ifnum [1]{%
 \ifnum #1\expandafter \@firstoftwo
 \else \expandafter \@secondoftwo
 \fi
}%
\providecommand \@ifx [1]{%
 \ifx #1\expandafter \@firstoftwo
 \else \expandafter \@secondoftwo
 \fi
}%
\providecommand \natexlab [1]{#1}%
\providecommand \enquote  [1]{``#1''}%
\providecommand \bibnamefont  [1]{#1}%
\providecommand \bibfnamefont [1]{#1}%
\providecommand \citenamefont [1]{#1}%
\providecommand \href@noop [0]{\@secondoftwo}%
\providecommand \href [0]{\begingroup \@sanitize@url \@href}%
\providecommand \@href[1]{\@@startlink{#1}\@@href}%
\providecommand \@@href[1]{\endgroup#1\@@endlink}%
\providecommand \@sanitize@url [0]{\catcode `\\12\catcode `\$12\catcode
  `\&12\catcode `\#12\catcode `\^12\catcode `\_12\catcode `\%12\relax}%
\providecommand \@@startlink[1]{}%
\providecommand \@@endlink[0]{}%
\providecommand \url  [0]{\begingroup\@sanitize@url \@url }%
\providecommand \@url [1]{\endgroup\@href {#1}{\urlprefix }}%
\providecommand \urlprefix  [0]{URL }%
\providecommand \Eprint [0]{\href }%
\providecommand \doibase [0]{http://dx.doi.org/}%
\providecommand \selectlanguage [0]{\@gobble}%
\providecommand \bibinfo  [0]{\@secondoftwo}%
\providecommand \bibfield  [0]{\@secondoftwo}%
\providecommand \translation [1]{[#1]}%
\providecommand \BibitemOpen [0]{}%
\providecommand \bibitemStop [0]{}%
\providecommand \bibitemNoStop [0]{.\EOS\space}%
\providecommand \EOS [0]{\spacefactor3000\relax}%
\providecommand \BibitemShut  [1]{\csname bibitem#1\endcsname}%
\let\auto@bib@innerbib\@empty
\bibitem [{\citenamefont {Larson}(1999)}]{Larson-book}%
  \BibitemOpen
  \bibfield  {author} {\bibinfo {author} {\bibfnamefont {R.~G.}\ \bibnamefont
  {Larson}},\ }\href@noop {} {\emph {\bibinfo {title} {The Structure and
  Rheology of Complex Fluids}}}\ (\bibinfo  {publisher} {Oxford University
  Press},\ \bibinfo {address} {New York},\ \bibinfo {year} {1999})\BibitemShut
  {NoStop}%
\bibitem [{\citenamefont {Annable}\ \emph {et~al.}(1993)\citenamefont
  {Annable}, \citenamefont {Buscall}, \citenamefont {Ettelaie},\ and\
  \citenamefont {Whittelestone}}]{Annable-Buscall-Ettelaie-Whittelstone-1993}%
  \BibitemOpen
  \bibfield  {author} {\bibinfo {author} {\bibfnamefont {T.}~\bibnamefont
  {Annable}}, \bibinfo {author} {\bibfnamefont {R.}~\bibnamefont {Buscall}},
  \bibinfo {author} {\bibfnamefont {R.}~\bibnamefont {Ettelaie}}, \ and\
  \bibinfo {author} {\bibfnamefont {D.}~\bibnamefont {Whittelestone}},\
  }\href@noop {} {\bibfield  {journal} {\bibinfo  {journal} {J. Rheol.}\
  }\textbf {\bibinfo {volume} {37}},\ \bibinfo {pages} {695} (\bibinfo {year}
  {1993})}\BibitemShut {NoStop}%
\bibitem [{\citenamefont {Tripathi}\ \emph {et~al.}(2006)\citenamefont
  {Tripathi}, \citenamefont {Tam},\ and\ \citenamefont
  {McKinley}}]{Tripathi-Tam-McKinley-2006}%
  \BibitemOpen
  \bibfield  {author} {\bibinfo {author} {\bibfnamefont {A.}~\bibnamefont
  {Tripathi}}, \bibinfo {author} {\bibfnamefont {K.~C.}\ \bibnamefont {Tam}}, \
  and\ \bibinfo {author} {\bibfnamefont {G.~H.}\ \bibnamefont {McKinley}},\
  }\href@noop {} {\bibfield  {journal} {\bibinfo  {journal} {Macromolecules}\
  }\textbf {\bibinfo {volume} {39}},\ \bibinfo {pages} {1981} (\bibinfo {year}
  {2006})}\BibitemShut {NoStop}%
\bibitem [{\citenamefont {Xu}\ \emph {et~al.}(1996)\citenamefont {Xu},
  \citenamefont {Yekta}, \citenamefont {Li}, \citenamefont {Masoumi},\ and\
  \citenamefont {Winnik}}]{Xu-Yekta-Li-Masoumi-Winnik-1996}%
  \BibitemOpen
  \bibfield  {author} {\bibinfo {author} {\bibfnamefont {B.}~\bibnamefont
  {Xu}}, \bibinfo {author} {\bibfnamefont {A.}~\bibnamefont {Yekta}}, \bibinfo
  {author} {\bibfnamefont {L.}~\bibnamefont {Li}}, \bibinfo {author}
  {\bibfnamefont {Z.}~\bibnamefont {Masoumi}}, \ and\ \bibinfo {author}
  {\bibfnamefont {M.~A.}\ \bibnamefont {Winnik}},\ }\href@noop {} {\bibfield
  {journal} {\bibinfo  {journal} {Colloilds Surf. A: Physicochem. Eng. Asp.}\
  }\textbf {\bibinfo {volume} {112}},\ \bibinfo {pages} {239} (\bibinfo {year}
  {1996})}\BibitemShut {NoStop}%
\bibitem [{\citenamefont {Tam}\ \emph {et~al.}(1998)\citenamefont {Tam},
  \citenamefont {Jenkins}, \citenamefont {Winnik},\ and\ \citenamefont
  {Bassett}}]{Tam-Jenkins-Winnik-Bassett-1998}%
  \BibitemOpen
  \bibfield  {author} {\bibinfo {author} {\bibfnamefont {K.~C.}\ \bibnamefont
  {Tam}}, \bibinfo {author} {\bibfnamefont {R.~D.}\ \bibnamefont {Jenkins}},
  \bibinfo {author} {\bibfnamefont {M.~A.}\ \bibnamefont {Winnik}}, \ and\
  \bibinfo {author} {\bibfnamefont {D.~R.}\ \bibnamefont {Bassett}},\
  }\href@noop {} {\bibfield  {journal} {\bibinfo  {journal} {Macromolecules}\
  }\textbf {\bibinfo {volume} {31}},\ \bibinfo {pages} {4149} (\bibinfo {year}
  {1998})}\BibitemShut {NoStop}%
\bibitem [{\citenamefont {Green}\ and\ \citenamefont
  {Tobolsky}(1946)}]{Green-Tobolsky-1946}%
  \BibitemOpen
  \bibfield  {author} {\bibinfo {author} {\bibfnamefont {M.~S.}\ \bibnamefont
  {Green}}\ and\ \bibinfo {author} {\bibfnamefont {A.~V.}\ \bibnamefont
  {Tobolsky}},\ }\href@noop {} {\bibfield  {journal} {\bibinfo  {journal} {J.
  Chem. Phys.}\ }\textbf {\bibinfo {volume} {14}},\ \bibinfo {pages} {80}
  (\bibinfo {year} {1946})}\BibitemShut {NoStop}%
\bibitem [{\citenamefont {Tanaka}\ and\ \citenamefont
  {Edwards}(1992)}]{Tanaka-Edwards-1992}%
  \BibitemOpen
  \bibfield  {author} {\bibinfo {author} {\bibfnamefont {F.}~\bibnamefont
  {Tanaka}}\ and\ \bibinfo {author} {\bibfnamefont {S.~F.}\ \bibnamefont
  {Edwards}},\ }\href@noop {} {\bibfield  {journal} {\bibinfo  {journal}
  {Macromolecules}\ }\textbf {\bibinfo {volume} {25}},\ \bibinfo {pages} {1516}
  (\bibinfo {year} {1992})}\BibitemShut {NoStop}%
\bibitem [{\citenamefont {Cath\'{e}bras}\ \emph {et~al.}(1998)\citenamefont
  {Cath\'{e}bras}, \citenamefont {Collet}, \citenamefont {Viguier},\ and\
  \citenamefont {Berret}}]{Cathebras-Collet-Viguier-Berret-1998}%
  \BibitemOpen
  \bibfield  {author} {\bibinfo {author} {\bibfnamefont {N.}~\bibnamefont
  {Cath\'{e}bras}}, \bibinfo {author} {\bibfnamefont {A.}~\bibnamefont
  {Collet}}, \bibinfo {author} {\bibfnamefont {M.}~\bibnamefont {Viguier}}, \
  and\ \bibinfo {author} {\bibfnamefont {J.-F.}\ \bibnamefont {Berret}},\
  }\href@noop {} {\bibfield  {journal} {\bibinfo  {journal} {Macromolecules}\
  }\textbf {\bibinfo {volume} {31}},\ \bibinfo {pages} {1305} (\bibinfo {year}
  {1998})}\BibitemShut {NoStop}%
\bibitem [{\citenamefont {S\'{e}r\'{e}ro}\ \emph {et~al.}(1998)\citenamefont
  {S\'{e}r\'{e}ro}, \citenamefont {Aznar}, \citenamefont {Porte}, \citenamefont
  {Berret}, \citenamefont {Calvet}, \citenamefont {Collet},\ and\ \citenamefont
  {Viguier}}]{Serero-Aznar-Porte-Berret-Calvet-Collet-Viguer-1998}%
  \BibitemOpen
  \bibfield  {author} {\bibinfo {author} {\bibfnamefont {Y.}~\bibnamefont
  {S\'{e}r\'{e}ro}}, \bibinfo {author} {\bibfnamefont {R.}~\bibnamefont
  {Aznar}}, \bibinfo {author} {\bibfnamefont {G.}~\bibnamefont {Porte}},
  \bibinfo {author} {\bibfnamefont {J.-F.}\ \bibnamefont {Berret}}, \bibinfo
  {author} {\bibfnamefont {D.}~\bibnamefont {Calvet}}, \bibinfo {author}
  {\bibfnamefont {A.}~\bibnamefont {Collet}}, \ and\ \bibinfo {author}
  {\bibfnamefont {M.}~\bibnamefont {Viguier}},\ }\href@noop {} {\bibfield
  {journal} {\bibinfo  {journal} {Phys. Rev. Lett.}\ }\textbf {\bibinfo
  {volume} {81}},\ \bibinfo {pages} {5584} (\bibinfo {year}
  {1998})}\BibitemShut {NoStop}%
\bibitem [{\citenamefont {S\'{e}r\'{e}ro}\ \emph {et~al.}(2000)\citenamefont
  {S\'{e}r\'{e}ro}, \citenamefont {Jacobsen},\ and\ \citenamefont
  {Berret}}]{Serero-Jacobsen-Berret-2000}%
  \BibitemOpen
  \bibfield  {author} {\bibinfo {author} {\bibfnamefont {Y.}~\bibnamefont
  {S\'{e}r\'{e}ro}}, \bibinfo {author} {\bibfnamefont {V.}~\bibnamefont
  {Jacobsen}}, \ and\ \bibinfo {author} {\bibfnamefont {J.-F.}\ \bibnamefont
  {Berret}},\ }\href@noop {} {\bibfield  {journal} {\bibinfo  {journal}
  {Macromolecules}\ }\textbf {\bibinfo {volume} {33}},\ \bibinfo {pages} {1841}
  (\bibinfo {year} {2000})}\BibitemShut {NoStop}%
\bibitem [{\citenamefont {Wang}(1992)}]{Wang-1992a}%
  \BibitemOpen
  \bibfield  {author} {\bibinfo {author} {\bibfnamefont {S.-Q.}\ \bibnamefont
  {Wang}},\ }\href@noop {} {\bibfield  {journal} {\bibinfo  {journal}
  {Macromolecules}\ }\textbf {\bibinfo {volume} {25}},\ \bibinfo {pages} {7003}
  (\bibinfo {year} {1992})}\BibitemShut {NoStop}%
\bibitem [{\citenamefont {Marrucci}\ \emph {et~al.}(1993)\citenamefont
  {Marrucci}, \citenamefont {Bhargava},\ and\ \citenamefont
  {Cooper}}]{Marrucci-Bhargava-Cooper-1933}%
  \BibitemOpen
  \bibfield  {author} {\bibinfo {author} {\bibfnamefont {G.}~\bibnamefont
  {Marrucci}}, \bibinfo {author} {\bibfnamefont {S.}~\bibnamefont {Bhargava}},
  \ and\ \bibinfo {author} {\bibfnamefont {S.~L.}\ \bibnamefont {Cooper}},\
  }\href@noop {} {\bibfield  {journal} {\bibinfo  {journal} {Macromolecules}\
  }\textbf {\bibinfo {volume} {26}},\ \bibinfo {pages} {6438} (\bibinfo {year}
  {1993})}\BibitemShut {NoStop}%
\bibitem [{\citenamefont {Vaccaro}\ and\ \citenamefont
  {Marrucci}(2000)}]{Vaccaro-Marrucci-2000}%
  \BibitemOpen
  \bibfield  {author} {\bibinfo {author} {\bibfnamefont {A.}~\bibnamefont
  {Vaccaro}}\ and\ \bibinfo {author} {\bibfnamefont {G.}~\bibnamefont
  {Marrucci}},\ }\href@noop {} {\bibfield  {journal} {\bibinfo  {journal} {J.
  Non-Newtonian Fluid Mech.}\ }\textbf {\bibinfo {volume} {92}},\ \bibinfo
  {pages} {261} (\bibinfo {year} {2000})}\BibitemShut {NoStop}%
\bibitem [{\citenamefont {Indei}\ \emph {et~al.}(2005)\citenamefont {Indei},
  \citenamefont {Koga},\ and\ \citenamefont {Tanaka}}]{Indei-Koga-Tanaka-2005}%
  \BibitemOpen
  \bibfield  {author} {\bibinfo {author} {\bibfnamefont {T.}~\bibnamefont
  {Indei}}, \bibinfo {author} {\bibfnamefont {T.}~\bibnamefont {Koga}}, \ and\
  \bibinfo {author} {\bibfnamefont {F.}~\bibnamefont {Tanaka}},\ }\href@noop {}
  {\bibfield  {journal} {\bibinfo  {journal} {Macromol. Rapid Comm.}\ }\textbf
  {\bibinfo {volume} {26}},\ \bibinfo {pages} {701} (\bibinfo {year}
  {2005})}\BibitemShut {NoStop}%
\bibitem [{\citenamefont {Indei}(2007{\natexlab{a}})}]{Indei-2007}%
  \BibitemOpen
  \bibfield  {author} {\bibinfo {author} {\bibfnamefont {T.}~\bibnamefont
  {Indei}},\ }\href@noop {} {\bibfield  {journal} {\bibinfo  {journal} {J.
  Non-Newtonian Fluid Mech.}\ }\textbf {\bibinfo {volume} {141}},\ \bibinfo
  {pages} {18} (\bibinfo {year} {2007}{\natexlab{a}})}\BibitemShut {NoStop}%
\bibitem [{\citenamefont {Indei}(2007{\natexlab{b}})}]{Indei-2007a}%
  \BibitemOpen
  \bibfield  {author} {\bibinfo {author} {\bibfnamefont {T.}~\bibnamefont
  {Indei}},\ }\href@noop {} {\bibfield  {journal} {\bibinfo  {journal} {Nihon
  Reoroji Gakkaishi (J. Soc. Rheol. Jpn.)}\ }\textbf {\bibinfo {volume} {35}},\
  \bibinfo {pages} {147} (\bibinfo {year} {2007}{\natexlab{b}})}\BibitemShut
  {NoStop}%
\bibitem [{\citenamefont {Koga}\ and\ \citenamefont
  {Tanaka}(2010)}]{Koga-Tanaka-2010}%
  \BibitemOpen
  \bibfield  {author} {\bibinfo {author} {\bibfnamefont {T.}~\bibnamefont
  {Koga}}\ and\ \bibinfo {author} {\bibfnamefont {F.}~\bibnamefont {Tanaka}},\
  }\href@noop {} {\bibfield  {journal} {\bibinfo  {journal} {Macromolecules}\
  }\textbf {\bibinfo {volume} {43}},\ \bibinfo {pages} {3052} (\bibinfo {year}
  {2010})}\BibitemShut {NoStop}%
\bibitem [{\citenamefont {Suzuki}\ \emph {et~al.}(2012)\citenamefont {Suzuki},
  \citenamefont {Uneyama}, \citenamefont {Inoue},\ and\ \citenamefont
  {Watanabe}}]{Suzuki-Uneyama-Inoue-Watanabe-2012}%
  \BibitemOpen
  \bibfield  {author} {\bibinfo {author} {\bibfnamefont {S.}~\bibnamefont
  {Suzuki}}, \bibinfo {author} {\bibfnamefont {T.}~\bibnamefont {Uneyama}},
  \bibinfo {author} {\bibfnamefont {T.}~\bibnamefont {Inoue}}, \ and\ \bibinfo
  {author} {\bibfnamefont {H.}~\bibnamefont {Watanabe}},\ }\href@noop {}
  {\bibfield  {journal} {\bibinfo  {journal} {Macromolecules}\ }\textbf
  {\bibinfo {volume} {45}},\ \bibinfo {pages} {888} (\bibinfo {year}
  {2012})}\BibitemShut {NoStop}%
\bibitem [{\citenamefont {Verduin}\ \emph {et~al.}(1996)\citenamefont
  {Verduin}, \citenamefont {de~Gans},\ and\ \citenamefont
  {Dhont}}]{Verduin-deGans-Dhont-1996}%
  \BibitemOpen
  \bibfield  {author} {\bibinfo {author} {\bibfnamefont {H.}~\bibnamefont
  {Verduin}}, \bibinfo {author} {\bibfnamefont {B.~J.}\ \bibnamefont
  {de~Gans}}, \ and\ \bibinfo {author} {\bibfnamefont {J.~K.~G.}\ \bibnamefont
  {Dhont}},\ }\href@noop {} {\bibfield  {journal} {\bibinfo  {journal}
  {Langmuir}\ }\textbf {\bibinfo {volume} {12}},\ \bibinfo {pages} {2947}
  (\bibinfo {year} {1996})}\BibitemShut {NoStop}%
\bibitem [{\citenamefont {Brader}(2010)}]{Brader-2010}%
  \BibitemOpen
  \bibfield  {author} {\bibinfo {author} {\bibfnamefont {J.~M.}\ \bibnamefont
  {Brader}},\ }\href@noop {} {\bibfield  {journal} {\bibinfo  {journal} {J.
  Phys.: Cond. Mat.}\ }\textbf {\bibinfo {volume} {22}},\ \bibinfo {pages}
  {363101} (\bibinfo {year} {2010})}\BibitemShut {NoStop}%
\bibitem [{\citenamefont {Semenov}\ \emph {et~al.}(1995)\citenamefont
  {Semenov}, \citenamefont {Joanny},\ and\ \citenamefont
  {Khokhlov}}]{Semenov-Joanny-Khokhlov-1995}%
  \BibitemOpen
  \bibfield  {author} {\bibinfo {author} {\bibfnamefont {A.~N.}\ \bibnamefont
  {Semenov}}, \bibinfo {author} {\bibfnamefont {J.-F.}\ \bibnamefont {Joanny}},
  \ and\ \bibinfo {author} {\bibfnamefont {A.~R.}\ \bibnamefont {Khokhlov}},\
  }\href@noop {} {\bibfield  {journal} {\bibinfo  {journal} {Macromolecules}\
  }\textbf {\bibinfo {volume} {28}},\ \bibinfo {pages} {1066} (\bibinfo {year}
  {1995})}\BibitemShut {NoStop}%
\bibitem [{\citenamefont {Khalatur}\ and\ \citenamefont
  {Khokhlov}(1996)}]{Khalatur-Khokhlov-1996}%
  \BibitemOpen
  \bibfield  {author} {\bibinfo {author} {\bibfnamefont {P.~G.}\ \bibnamefont
  {Khalatur}}\ and\ \bibinfo {author} {\bibfnamefont {A.~R.}\ \bibnamefont
  {Khokhlov}},\ }\href@noop {} {\bibfield  {journal} {\bibinfo  {journal}
  {Macromol. Theory Simul.}\ }\textbf {\bibinfo {volume} {5}},\ \bibinfo
  {pages} {877} (\bibinfo {year} {1996})}\BibitemShut {NoStop}%
\bibitem [{\citenamefont {Ng}\ \emph {et~al.}(2000)\citenamefont {Ng},
  \citenamefont {Tam},\ and\ \citenamefont {Jenkins}}]{Ng-Tam-Jenkins-2000}%
  \BibitemOpen
  \bibfield  {author} {\bibinfo {author} {\bibfnamefont {W.~K.}\ \bibnamefont
  {Ng}}, \bibinfo {author} {\bibfnamefont {K.~C.}\ \bibnamefont {Tam}}, \ and\
  \bibinfo {author} {\bibfnamefont {R.~D.}\ \bibnamefont {Jenkins}},\
  }\href@noop {} {\bibfield  {journal} {\bibinfo  {journal} {J. Rheol.}\
  }\textbf {\bibinfo {volume} {44}},\ \bibinfo {pages} {137} (\bibinfo {year}
  {2000})}\BibitemShut {NoStop}%
\bibitem [{\citenamefont {Bedrov}\ \emph {et~al.}(2002)\citenamefont {Bedrov},
  \citenamefont {Smith},\ and\ \citenamefont
  {Douglas}}]{Bedrov-Smith-Douglas-2002}%
  \BibitemOpen
  \bibfield  {author} {\bibinfo {author} {\bibfnamefont {D.}~\bibnamefont
  {Bedrov}}, \bibinfo {author} {\bibfnamefont {G.~D.}\ \bibnamefont {Smith}}, \
  and\ \bibinfo {author} {\bibfnamefont {J.~F.}\ \bibnamefont {Douglas}},\
  }\href@noop {} {\bibfield  {journal} {\bibinfo  {journal} {Europhys. Lett.}\
  }\textbf {\bibinfo {volume} {59}},\ \bibinfo {pages} {384} (\bibinfo {year}
  {2002})}\BibitemShut {NoStop}%
\bibitem [{\citenamefont {Tae}\ \emph {et~al.}(2002)\citenamefont {Tae},
  \citenamefont {Kornfield}, \citenamefont {Hubbell},\ and\ \citenamefont
  {Lal}}]{Tae-Kornfield-Hubbell-Lal-2002}%
  \BibitemOpen
  \bibfield  {author} {\bibinfo {author} {\bibfnamefont {G.}~\bibnamefont
  {Tae}}, \bibinfo {author} {\bibfnamefont {J.~A.}\ \bibnamefont {Kornfield}},
  \bibinfo {author} {\bibfnamefont {J.~A.}\ \bibnamefont {Hubbell}}, \ and\
  \bibinfo {author} {\bibfnamefont {J.}~\bibnamefont {Lal}},\ }\href@noop {}
  {\bibfield  {journal} {\bibinfo  {journal} {Macromolecules}\ }\textbf
  {\bibinfo {volume} {35}},\ \bibinfo {pages} {4448} (\bibinfo {year}
  {2002})}\BibitemShut {NoStop}%
\bibitem [{\citenamefont {Berret}\ \emph {et~al.}(2003)\citenamefont {Berret},
  \citenamefont {Calvet}, \citenamefont {Collet},\ and\ \citenamefont
  {Viguier}}]{Berret-Calvet-Collet-Viguier-2003}%
  \BibitemOpen
  \bibfield  {author} {\bibinfo {author} {\bibfnamefont {J.-F.}\ \bibnamefont
  {Berret}}, \bibinfo {author} {\bibfnamefont {D.}~\bibnamefont {Calvet}},
  \bibinfo {author} {\bibfnamefont {A.}~\bibnamefont {Collet}}, \ and\ \bibinfo
  {author} {\bibfnamefont {M.}~\bibnamefont {Viguier}},\ }\href@noop {}
  {\bibfield  {journal} {\bibinfo  {journal} {Curr. Opin. Colloid Interface
  Sci.}\ }\textbf {\bibinfo {volume} {8}},\ \bibinfo {pages} {296} (\bibinfo
  {year} {2003})}\BibitemShut {NoStop}%
\bibitem [{\citenamefont {Castelletto}\ \emph {et~al.}(2004)\citenamefont
  {Castelletto}, \citenamefont {Hamley}, \citenamefont {Xue}, \citenamefont
  {Sommer}, \citenamefont {Pedersen},\ and\ \citenamefont
  {Olmsted}}]{Castelleto-Hamley-Xue-Sommer-Pedersen-Olmsted-2004}%
  \BibitemOpen
  \bibfield  {author} {\bibinfo {author} {\bibfnamefont {V.}~\bibnamefont
  {Castelletto}}, \bibinfo {author} {\bibfnamefont {I.~W.}\ \bibnamefont
  {Hamley}}, \bibinfo {author} {\bibfnamefont {W.}~\bibnamefont {Xue}},
  \bibinfo {author} {\bibfnamefont {C.}~\bibnamefont {Sommer}}, \bibinfo
  {author} {\bibfnamefont {J.~S.}\ \bibnamefont {Pedersen}}, \ and\ \bibinfo
  {author} {\bibfnamefont {P.~D.}\ \bibnamefont {Olmsted}},\ }\href@noop {}
  {\bibfield  {journal} {\bibinfo  {journal} {Macromolecules}\ }\textbf
  {\bibinfo {volume} {37}},\ \bibinfo {pages} {1492} (\bibinfo {year}
  {2004})}\BibitemShut {NoStop}%
\bibitem [{\citenamefont {Cass}\ \emph {et~al.}(2008)\citenamefont {Cass},
  \citenamefont {Heyes}, \citenamefont {Blanchard},\ and\ \citenamefont
  {English}}]{Cass-Heyes-Blanchard-English-2008}%
  \BibitemOpen
  \bibfield  {author} {\bibinfo {author} {\bibfnamefont {M.~J.}\ \bibnamefont
  {Cass}}, \bibinfo {author} {\bibfnamefont {D.~M.}\ \bibnamefont {Heyes}},
  \bibinfo {author} {\bibfnamefont {R.-L.}\ \bibnamefont {Blanchard}}, \ and\
  \bibinfo {author} {\bibfnamefont {R.~J.}\ \bibnamefont {English}},\
  }\href@noop {} {\bibfield  {journal} {\bibinfo  {journal} {J. Phys.: Cond.
  Mat.}\ }\textbf {\bibinfo {volume} {20}},\ \bibinfo {pages} {335103}
  (\bibinfo {year} {2008})}\BibitemShut {NoStop}%
\bibitem [{\citenamefont {Sprakel}\ \emph {et~al.}(2009)\citenamefont
  {Sprakel}, \citenamefont {Spruijt}, \citenamefont {van~der Gucht},
  \citenamefont {Padding},\ and\ \citenamefont
  {Briels}}]{Sprakel-Spruijt-vanderGucht-Padding-Briels-2009}%
  \BibitemOpen
  \bibfield  {author} {\bibinfo {author} {\bibfnamefont {J.}~\bibnamefont
  {Sprakel}}, \bibinfo {author} {\bibfnamefont {E.}~\bibnamefont {Spruijt}},
  \bibinfo {author} {\bibfnamefont {J.}~\bibnamefont {van~der Gucht}}, \bibinfo
  {author} {\bibfnamefont {J.~T.}\ \bibnamefont {Padding}}, \ and\ \bibinfo
  {author} {\bibfnamefont {W.~J.}\ \bibnamefont {Briels}},\ }\href@noop {}
  {\bibfield  {journal} {\bibinfo  {journal} {Soft Matter}\ }\textbf {\bibinfo
  {volume} {5}},\ \bibinfo {pages} {4748} (\bibinfo {year} {2009})}\BibitemShut
  {NoStop}%
\bibitem [{\citenamefont {Mistry}\ \emph {et~al.}(2006)\citenamefont {Mistry},
  \citenamefont {Annable}, \citenamefont {Yuan},\ and\ \citenamefont
  {Booth}}]{Mistry-Annable-Yuan-Booth-2006}%
  \BibitemOpen
  \bibfield  {author} {\bibinfo {author} {\bibfnamefont {D.}~\bibnamefont
  {Mistry}}, \bibinfo {author} {\bibfnamefont {T.}~\bibnamefont {Annable}},
  \bibinfo {author} {\bibfnamefont {X.-F.}\ \bibnamefont {Yuan}}, \ and\
  \bibinfo {author} {\bibfnamefont {C.}~\bibnamefont {Booth}},\ }\href@noop {}
  {\bibfield  {journal} {\bibinfo  {journal} {Langmuir}\ }\textbf {\bibinfo
  {volume} {22}},\ \bibinfo {pages} {2986} (\bibinfo {year}
  {2006})}\BibitemShut {NoStop}%
\bibitem [{\citenamefont {Hansen}\ and\ \citenamefont
  {McDonald}(2006)}]{Hansen-McDonald-book}%
  \BibitemOpen
  \bibfield  {author} {\bibinfo {author} {\bibfnamefont {J.-P.}\ \bibnamefont
  {Hansen}}\ and\ \bibinfo {author} {\bibfnamefont {I.~R.}\ \bibnamefont
  {McDonald}},\ }\href@noop {} {\emph {\bibinfo {title} {Theory of Simple
  Liquids}}},\ \bibinfo {edition} {3rd}\ ed.\ (\bibinfo  {publisher}
  {Elsevier},\ \bibinfo {address} {Amsterdam},\ \bibinfo {year}
  {2006})\BibitemShut {NoStop}%
\bibitem [{\citenamefont {van Kampen}(2007)}]{vanKampen-book}%
  \BibitemOpen
  \bibfield  {author} {\bibinfo {author} {\bibfnamefont {N.~G.}\ \bibnamefont
  {van Kampen}},\ }\href@noop {} {\emph {\bibinfo {title} {Stochastic Processes
  in Physics and Chemistry}}},\ \bibinfo {edition} {3rd}\ ed.\ (\bibinfo
  {publisher} {Elsevier},\ \bibinfo {address} {Amsterdam},\ \bibinfo {year}
  {2007})\BibitemShut {NoStop}%
\bibitem [{\citenamefont {McPhie}\ \emph {et~al.}(2001)\citenamefont {McPhie},
  \citenamefont {Daivis}, \citenamefont {Snook}, \citenamefont {Ennis},\ and\
  \citenamefont {Evans}}]{McPhie-Daivis-Snook-Ennis-Evans-2001}%
  \BibitemOpen
  \bibfield  {author} {\bibinfo {author} {\bibfnamefont {M.~G.}\ \bibnamefont
  {McPhie}}, \bibinfo {author} {\bibfnamefont {P.~J.}\ \bibnamefont {Daivis}},
  \bibinfo {author} {\bibfnamefont {I.~K.}\ \bibnamefont {Snook}}, \bibinfo
  {author} {\bibfnamefont {J.}~\bibnamefont {Ennis}}, \ and\ \bibinfo {author}
  {\bibfnamefont {D.~J.}\ \bibnamefont {Evans}},\ }\href@noop {} {\bibfield
  {journal} {\bibinfo  {journal} {Physica A}\ }\textbf {\bibinfo {volume}
  {299}},\ \bibinfo {pages} {412} (\bibinfo {year} {2001})}\BibitemShut
  {NoStop}%
\bibitem [{\citenamefont {Uneyama}\ \emph {et~al.}(2011)\citenamefont
  {Uneyama}, \citenamefont {Horio},\ and\ \citenamefont
  {Watanabe}}]{Uneyama-Horio-Watanabe-2011}%
  \BibitemOpen
  \bibfield  {author} {\bibinfo {author} {\bibfnamefont {T.}~\bibnamefont
  {Uneyama}}, \bibinfo {author} {\bibfnamefont {K.}~\bibnamefont {Horio}}, \
  and\ \bibinfo {author} {\bibfnamefont {H.}~\bibnamefont {Watanabe}},\
  }\href@noop {} {\bibfield  {journal} {\bibinfo  {journal} {Phys. Rev. E}\
  }\textbf {\bibinfo {volume} {83}},\ \bibinfo {pages} {061802} (\bibinfo
  {year} {2011})}\BibitemShut {NoStop}%
\bibitem [{\citenamefont {Evans}\ and\ \citenamefont
  {Morris}(2008)}]{Evans-Morris-book}%
  \BibitemOpen
  \bibfield  {author} {\bibinfo {author} {\bibfnamefont {D.~J.}\ \bibnamefont
  {Evans}}\ and\ \bibinfo {author} {\bibfnamefont {G.~P.}\ \bibnamefont
  {Morris}},\ }\href@noop {} {\emph {\bibinfo {title} {Statistical Mechanics of
  Nonequilibrium Liquids}}},\ \bibinfo {edition} {2nd}\ ed.\ (\bibinfo
  {publisher} {Cambridge University Press},\ \bibinfo {address} {Cambridge},\
  \bibinfo {year} {2008})\BibitemShut {NoStop}%
\bibitem [{\citenamefont {Suzuki}\ \emph {et~al.}()\citenamefont {Suzuki} \emph
  {et~al.}}]{Suzuki-inpreparation}%
  \BibitemOpen
  \bibfield  {author} {\bibinfo {author} {\bibfnamefont {S.}~\bibnamefont
  {Suzuki}} \emph {et~al.},\ }\href@noop {} {}\bibinfo {howpublished} {in
  preparation}\BibitemShut {NoStop}%
\bibitem [{\citenamefont {Abramowitz}\ and\ \citenamefont
  {Stegun}(1972)}]{Abramowitz-Stegun-book}%
  \BibitemOpen
  \bibinfo {editor} {\bibfnamefont {M.}~\bibnamefont {Abramowitz}}\ and\
  \bibinfo {editor} {\bibfnamefont {I.~A.}\ \bibnamefont {Stegun}},\ eds.,\
  \href@noop {} {\emph {\bibinfo {title} {Handbook of Mathematical Functions
  with Formulas, Graphs, and Mathematical Tables}}},\ \bibinfo {edition}
  {10th}\ ed.\ (\bibinfo  {publisher} {Dover},\ \bibinfo {address} {New York},\
  \bibinfo {year} {1972})\BibitemShut {NoStop}%
\bibitem [{\citenamefont {Tanaka}\ and\ \citenamefont
  {Koga}(2006)}]{Tanaka-Koga-2006}%
  \BibitemOpen
  \bibfield  {author} {\bibinfo {author} {\bibfnamefont {F.}~\bibnamefont
  {Tanaka}}\ and\ \bibinfo {author} {\bibfnamefont {T.}~\bibnamefont {Koga}},\
  }\href@noop {} {\bibfield  {journal} {\bibinfo  {journal} {Macromolecules}\
  }\textbf {\bibinfo {volume} {39}},\ \bibinfo {pages} {5913} (\bibinfo {year}
  {2006})}\BibitemShut {NoStop}%
\bibitem [{\citenamefont {Winter}\ and\ \citenamefont
  {Chambon}(1986)}]{Winter-Chambon-1986}%
  \BibitemOpen
  \bibfield  {author} {\bibinfo {author} {\bibfnamefont {H.~H.}\ \bibnamefont
  {Winter}}\ and\ \bibinfo {author} {\bibfnamefont {F.}~\bibnamefont
  {Chambon}},\ }\href@noop {} {\bibfield  {journal} {\bibinfo  {journal} {J.
  Rheol.}\ }\textbf {\bibinfo {volume} {30}},\ \bibinfo {pages} {367} (\bibinfo
  {year} {1986})}\BibitemShut {NoStop}%
\end{thebibliography}%

\clearpage

\section*{Figure Captions}

\begin{figure}[h]
 \centering
 {\includegraphics[width=0.9\linewidth,clip]{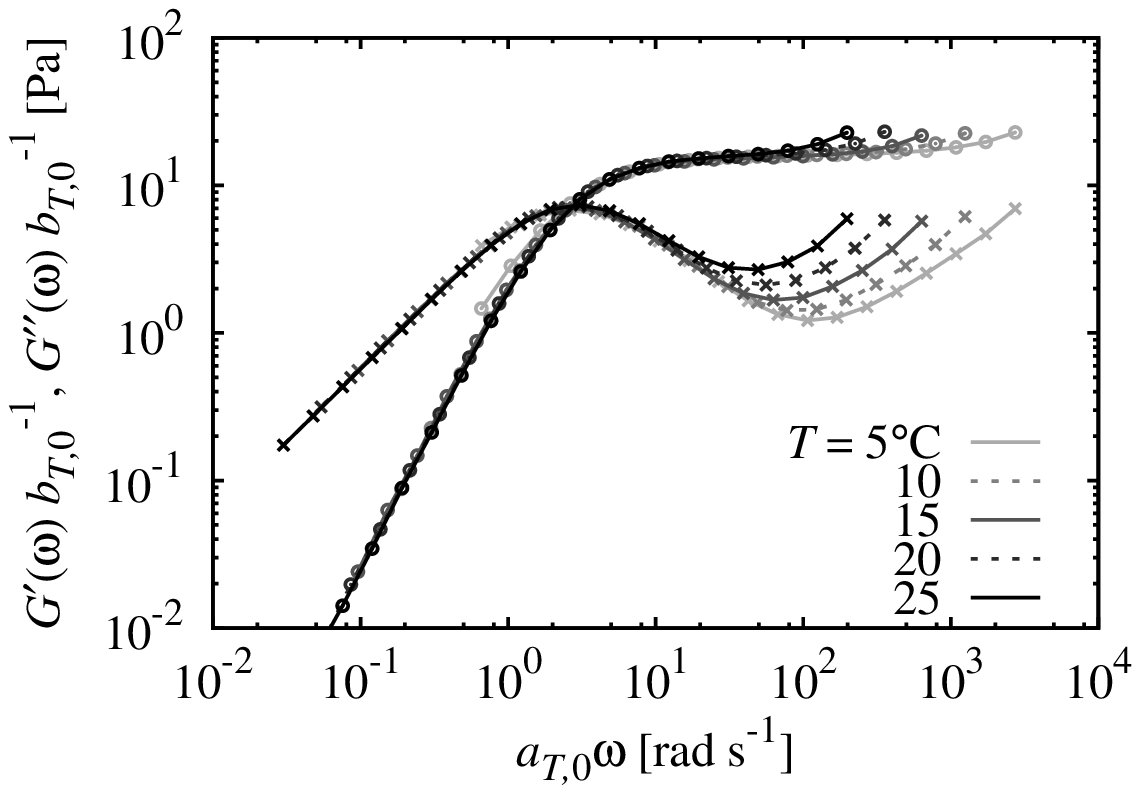}}
 \caption{\label{heur_1wt_master_curves_slow}}
\end{figure}

Figure \ref{heur_1wt_master_curves_slow}:
Master curves of storage and loss moduli for the $1$wt\% HEUR aqueous
solution. The reference temperature is $T_{r} = 25^{\circ}\text{C}$ and
the superposition is performed for the slow (low frequency) mode.

\

Figure \ref{heur_1wt_master_curves_fast}:
Master curves of storage and loss moduli for the $1$wt\% HEUR aqueous
solution. Data are the same as in Figure \ref{heur_1wt_master_curves_slow}
but shifted with $a_{T,1}$ determined for the $1\text{wt\%}$ PEO solution.
The reference temperature is
$T_{r} = 25^{\circ}\text{C}$.
For comparison, the loss moduli for the $1$wt\% PEO
solution at $T = 25^{\circ}\text{C}$ is also plotted.

\

Figure \ref{heur_1wt_shift_factor}:
Horizontal shift factors of the slow mode of the HEUR solution and the fast
mode of the PEO solution, $a_{T,0}$ and
$a_{T,1}$. The reference temperature is
$T_{r} = 25^{\circ}\text{C}$.

\

Figure \ref{heur_concentration_dependence}:
Concentration dependence of storage and loss moduli for the HEUR aqueous
solutions at $T = 25^{\circ}\text{C}$. The concentrations are $c = 1, 2, 5,$ and $10\text{wt\%}$.

\

Figure \ref{heur_concentration_dependence_rescaled}:
Storage and loss moduli for the HEUR aqueous
solutions with various concentrations at $T =
25^{\circ}\text{C}$. Angular frequency and moduli are rescaled by the
inverse characteristic time $1 / \tau$ and the characteristic
modulus $G_{0}$ for the slow mode, respectively.

\

Figure \ref{heur_tau0_g0_concentration_dependence}:
Concentration dependence of (a) the characteristic time $\tau$ and
(b) the characteristic modulus $G_{0}$ at $T
= 25^{\circ}\text{C}$ obtained for the HEUR aqueous solutions.
Broken and solid lines represent the fitting
results for low and high concentrations, respectively.

\

Figure \ref{schematic_image_of_single_chain_model}: A schematic image of
the mean field single
chain model for a dense network (the dense network model). (a) a dense network formed by
elastically active chains (solid black curves) and micellar cores
(circles). (b) a local structure in a dense network.
Solid and dotted curves represent elastically active
and inactive polymer chains, respectively.
Circles represent micellar cores. The dense network model is
formulated for a single tagged chain in the system.
There is the effective interaction between micelles, which is expressed
by the potential $v(\bm{r})$. A bridge chain feels the elastic potential $u(\bm{r})$.

\

\hspace{-\parindent}%
Figure \ref{schematic_image_of_single_effective_bond_model}: A schematic
image of the mean field single
chain model for a sparse network (the sparse network model).
(a) a sparse network formed by elastically active chains (solid black
curves) and micellar cores (circles). (b) a local structure in a sparse network.
Solid and dotted curves represent elastically active
and inactive polymer chains, respectively.
Circles represent micellar cores.
The sparse network model is
formulated for a single tagged superbridge in the system.
$v(\bm{r})$ and $\bar{u}(\bm{r})$ are the effective interaction
potential between micelles and the elastic potential of a superbridge.

\

\hspace{-\parindent}%
Figure \ref{storage_and_loss_moduli_dense_network}: Storage and loss
moduli ($G'(\omega)$ and $G''(\omega)$) deduced from a dense network
model for $\tilde{\tau} = \tilde{\tau}(0) = \tilde{\tau}(1)$ calculated
under some approximations. For comparison, $G'(\omega)$ and $G''(\omega)$ of a
single Maxwellian model ($G_{1} / G_{0} = 0$) are also
shown.

\clearpage

\section*{Figures}

\end{document}